\pdfoutput=1
\documentclass[12pt,onecolumn,oneside,a4paper,peerreview]{IEEEtran}
\usepackage{cite}
\usepackage{float}
\usepackage{graphicx}
\usepackage{amsmath}
\usepackage{times}
\usepackage{graphicx}
\usepackage{latexsym}
\usepackage{graphicx}
\usepackage{bm}
\usepackage{amssymb}
\usepackage[center]{caption2}
\usepackage{stfloats}
\usepackage{cases}
\usepackage{array}
\usepackage{setspace}
\usepackage{fancyhdr}
\usepackage{color}
\usepackage{subfigure}
\usepackage{citesort}
\usepackage{multirow}
\usepackage{amsthm}
\usepackage{cases}
\usepackage[noend]{algpseudocode}
\usepackage{algorithmicx,algorithm}
 \usepackage[long]{optidef}
\def\bm#1{\mbox{\boldmath $#1$}}

\newtheorem{remark}{Remark}
\renewcommand{\baselinestretch}{0.97}

\algdef{SE}[DOWHILE]{Do}{doWhile}{\algorithmicdo}[1]{\algorithmicwhile\ #1}

\begin{document}
\title{\color{black} IRS-Based Integrated Location Sensing and Communication for mmWave SIMO Systems}
\author{\IEEEauthorblockN{Xiaoling Hu, {\em Member, IEEE},   Chenxi Liu, {\em Member, IEEE},\\ Mugen Peng, {\em Fellow, IEEE}  and Caijun Zhong, {\em Senior Member, IEEE}}
\thanks{Xiaoling Hu, Chenxi Liu  and Mugen Peng  are with the State Key Laboratory of Networking and Switching Technology, Beijing University of Posts and
Telecommunications, Beijing 100876, China (e-mail: xiaolinghu@bupt.edu.cn; chenxi.liu@bupt.edu.cn;
pmg@bupt.edu.cn).}
\thanks{Caijun Zhong is with the College of Information Science and Electronic Engineering, Zhejiang University, Hangzhou 310027, China.(email: caijunzhong@zju.edu.cn).}
}
\maketitle
\begin{abstract}
    In this paper, we establish an integrated sensing and communication (ISAC) system based on a distributed semi-passive intelligent reflecting surface (IRS), which allows location sensing and data transmission to be carried out simultaneously, sharing the same frequency and time resources. The detailed working process of the proposed IRS-based ISAC system is designed, including the transmission protocol, location sensing and beamforming optimization. Specifically, each coherence block consists of two periods, the ISAC period with two time blocks and the pure communication (PC) period. During each time block of the ISAC period, data transmission and user positioning are carried out simultaneously. The estimated user location in the first time block will be used for beamforming design in the second time block. During the PC period, only data transmission is conducted, by invoking the user location estimated in the second time block of the ISAC period for beamforming design. {\color{black}Simulation results show that a  millimeter-level  positioning accuracy can be achieved by the proposed location sensing scheme,  demonstrating the advantage of the proposed IRS-based ISAC framework. Besides, the proposed two beamforming schemes  based on the estimated location information achieve similar performance to the benchmark schemes assuming perfect channel state information (CSI), which verifies the effectiveness of beamforming design using sensed location information.}
\end{abstract}
\begin{IEEEkeywords}
Integrated sensing and communication (ISAC), intelligent reflecting surface (IRS), beamforming design, and location sensing.
\end{IEEEkeywords}

\section{introduction}
Recently, intelligent reflecting surface (IRS) has been proposed as a promising candidate technology for the sixth-generation (6G) wireless system, due to its capability of realizing programmable propagation environment with low power consumption and hardware cost. Specifically, the IRS is a two-dimensional (2D) meta-surface composed of a large number of low-cost reflecting elements, each of which can independently  reflect the impinging signals with adjustable phase shits. The adjustment of phase shifts can be realized by cheap positive intrinsic negative (PIN) diodes or varactor diodes \cite{hardware}. Moreover, the IRS can be flexibly deployed  to assist data transmission without equipping any power-consuming transmit radio-frequency
(RF) chains \cite{survey1}.  Due to the above advantages, the IRS has attracted considerable research interests.

In particular, there are massive research works adopting the IRS to enhance communication performance. By properly deploying the IRS between the transmitter and the receiver, a strong virtual line-of-sight (VLoS) link between them can be  established. Through smartly designing IRS phase shifts, significant performance improvement can be achieved. For example, the work\cite{qingqing1} has demonstrated that a power gain of $M^2$ can be obtained by applying the IRS with $M$ reflecting elements. Also, the work \cite{9502509} has shown that applying the IRS with a large number of reflecting elements  helps reduce the outage probability.
 {\color{black}Moreover, the works \cite{BX1,BX6} proposed a double-IRS assisted system, and it was proved that the double-IRS assisted system with cooperative passive beamforming design is superior to the conventional single-IRS system in terms of both the maximum signal-to-noise ratio (SNR) and the multi-user effective channel rank.} Due to its great benefit in improving communication performance, the IRS has been extensively used in various communication scenarios to realize diverse desired goals, such as  SNR or capacity maximization\cite{location_hu,BX4}, {\color{black}sum rate maximization\cite{pan1}}, power minimization or energy efficiency  maximization\cite{qingqing1,9408385}, and symbol-error-rate  minimization \cite{9097454,8928065}. Also, the IRS has been proposed to be integrated with other promising technologies, such as {\color{black}orthogonal frequency division multiplexing (OFDM) \cite{BX2,BX3}}, massive MIMO\cite{9528043,BX5}, millimeter wave (mmWave)\cite{9226616,9410435},  deep learning\cite{9505267,9264659},  cognitive  radio\cite{9235486,9146170},  physical layer security\cite{9428001,9446526}, unmanned aerial vehicle  (UAV)\cite{9400768,9234511}, and {\color{black}simultaneous wireless information and power transfer (SWIPT)\cite{pan2}}, to improve the communication performance of the considered systems.

In addition to improving the communication performance, another important promising function of the  IRS is to assist location sensing in the wireless communication system.
In general, the mobile user can be localized according to channel parameters, such as, time of arrival (TOA)/time of difference arrival (TDOA) \cite{location_wireless_TOA}, angle of arrival (AOA)\cite{location_mmWave_AOA1} and  received signal strength (RSS)\cite{location_cellular_RSS1}. However, the RSS-based localization  has a poor location accuracy, which is affected by the network topology and propagation environment, e.g., path loss exponent and shadowing effects.
Although the TOA/TDOA-based and AOA-based localization can achieve a high location accuracy, they rely heavily on the line-of-sight (LoS) link, which may be blocked especially in the mmWave case.
Responding to this, the IRS has been proposed to overcome the blockage problem and improve the  location accuracy in the wireless communication system, due to its capability of establishing a strong VLoS path between the BS and the mobile user. As an early research, the work \cite{location_LIS} first explored the potential of  using IRS for wireless localization, where the Cramer-Rao lower bounds (CRLB) for positioning with IRS has been derived. Later on, J. He, {\em et al.} \cite{location_LIS_mmWave} investigated the 2D mmWave positioning with the assistance of a IRS, where the impact of the IRS on the location accuracy was evaluated. Then, the IRS-aided 2D mmWave positioning was extended to three-dimensional (3D) positioning\cite{location_RIS_3D}. Furthermore, an AoA-based mmWave positioning algorithm with the assistance of the IRS was proposed in \cite{location_IRS_mmWave2}, which achieves centimeter-level positioning accuracy.
H. Zhang, {\em et al.} \cite{location_RIS_RSS,location_RIS_RSS} considered the RSS-based localization, where the IRS was used to enlarge differences between the RSS values of adjacent locations so that a higher positioning accuracy can be achieved. In turn, the user location, provided by the global positioning system (GPS), was used for the design of IRS phase shifts in \cite{location_hu}.

In all the aforementioned works, location sensing and communication systems with the assistance of the IRS are usually designed separately and occupy different spectrum resources. With the wide deployment of the mmWave and massive multiple-input multiple-output (MIMO) technologies, it is possible to realize high-accuracy location sensing using communication signals. As such, it is desirable to jointly design the sensing and communication systems such that they can share the same frequency and time resources to improve the spectrum efficiency. This motivates the IRS-based integrated
sensing and communication (ISAC) system, where the IRS is introduced into the ISAC system to overcome the blockage problem in the mmWave system and maintain/improve both the communication performance and the sensing accuracy.  However, there are few works investigating the design of the IRS-based ISAC system.
A recent work related to the concept of IRS-based ISAC was proposed in \cite{IRS_JLC}, where a specific framework of the joint location and communication (JLC) system was designed. Since the location sensing and data transmission cannot  be conducted  at the same time (i.e., share the same time resources),  the IRS-aided JLC system cannot be counted as a real ISAC system in a strict sense, which allows  sensing and communication to share the same time and frequency resources.

In this paper, we establish an ISAC system realized by a distributed semi-passive IRS, and  design a specific framework of its working process, including transmission protocol design, location sensing and beamforming design. To the best of our knowledge, this is the first work investigating the IRS-based ISAC system, where location sensing and data transmission are conducted simultaneously, sharing the same time and frequency resources. Our main contributions are summarized as follows.
\begin{itemize}
    \item We construct a 3D ISAC system realized by a novel IRS architecture, i.e., the  distributed semi-passive IRS architecture. In the IRS-based ISAC system, location sensing is performed at the IRS by using  communication signals, and the obtained location information is used for beamforming design such that the communication performance is improved.
    \item A transmission protocol is proposed for the IRS-based ISAC system.  Specifically,  a coherence block consists of two periods, i.e., ISAC period and pure communication (PC) period. During the ISAC period, the mobile user sends information-carrying signals to the BS. Two semi-passive sub-IRSs operate in the sensing mode for localizing the user, while the passive sub-IRS operates in the sensing mode for assisting data transmission. The ISAC period consists of two time blocks. The location estimated in the first time block will be used for beamforming design in the second time block. During the PC period, two semi-passive sub-IRSs switch into the reflecting mode for data transmission, and the user location estimated in the second time block of the ISAC period is used for beamforming design so that communication performance can be improved.
    \item  {\color{black} We propose a location sensing scheme to localize the mobile user at the IRS, by using communication signals, thereby removing the requirement of dedicated positioning reference signals in conventional localization  methods. Simulation results demonstrate that  a millimeter-level positioning accuracy can be achieved.}
    \item We propose two location-based beamforming schemes for the ISAC and PC periods, respectively, where both the BS combining vector and the IRS phase shift beam are designed according to the estimated user location. Simulation results show that although only imperfect location information is available, the proposed beamforming scheme for the ISAC period has almost the same performance as the optimal beamforming scheme with perfect CSI, and the proposed beamforming scheme for the PC period achieves similar performance to the alternating optimization (AO) beamforming scheme
   assuming perfect CSI. {\color{black}These observations demonstrate that using sensed location information for beamforming can ensure communication performance.}
\end{itemize}

The remainder of the paper is organized as follows. In Section \ref{s1}, we introduce the IRS-based ISAC system, while in Section \ref{s2}, we propose a location sensing scheme. According to the estimated  user location,  low-complexity schemes for BS beamforming and IRS beamforming are presented in Section \ref{s3}.
Numerical results and discussions are provided in Section \ref{s4}, and finally Section \ref{s5} concludes the paper.

Notation: Boldface lower case and upper case letters are used for column vectors and matrices, respectively. The superscripts ${\left(\cdot\right)}^{*}$, ${\left(\cdot\right)}^{T}$, ${\left(\cdot\right)}^{H}$, and ${\left(\cdot\right)}^{-1}$ stand for the conjugate, transpose, conjugate-transpose, and matrix inverse, respectively. Also, the Euclidean norm, absolute value, Kronecker product  are denoted by $\left\| \cdot \right\|$, $\left|\cdot\right|$ and $\otimes$ respectively. In addition, $\mathbb{E}\left\{\cdot\right\}$ is the expectation operator, and $\text{tr}\left(\cdot\right)$ represents the trace.
For a matrix ${\bf A}$, ${[\bf A]}_{mn}$ denotes its entry in the $m$-th row and $n$-th column, while for a vector ${\bf a}$, ${[\bf a]}_{m}$ denotes the $m$-th entry of it. Besides, $j$ in $e^{j \theta}$ denotes the imaginary unit.
Finally, $z \sim \mathcal{CN}(0,{\sigma}^{2})$ denotes a circularly symmetric complex Gaussian random variable $z$ with zero mean and variance $\sigma^2$.


\section{System Model} \label{s1}
\begin{figure}[!ht]
  \centering
  \includegraphics[width=4.5in]{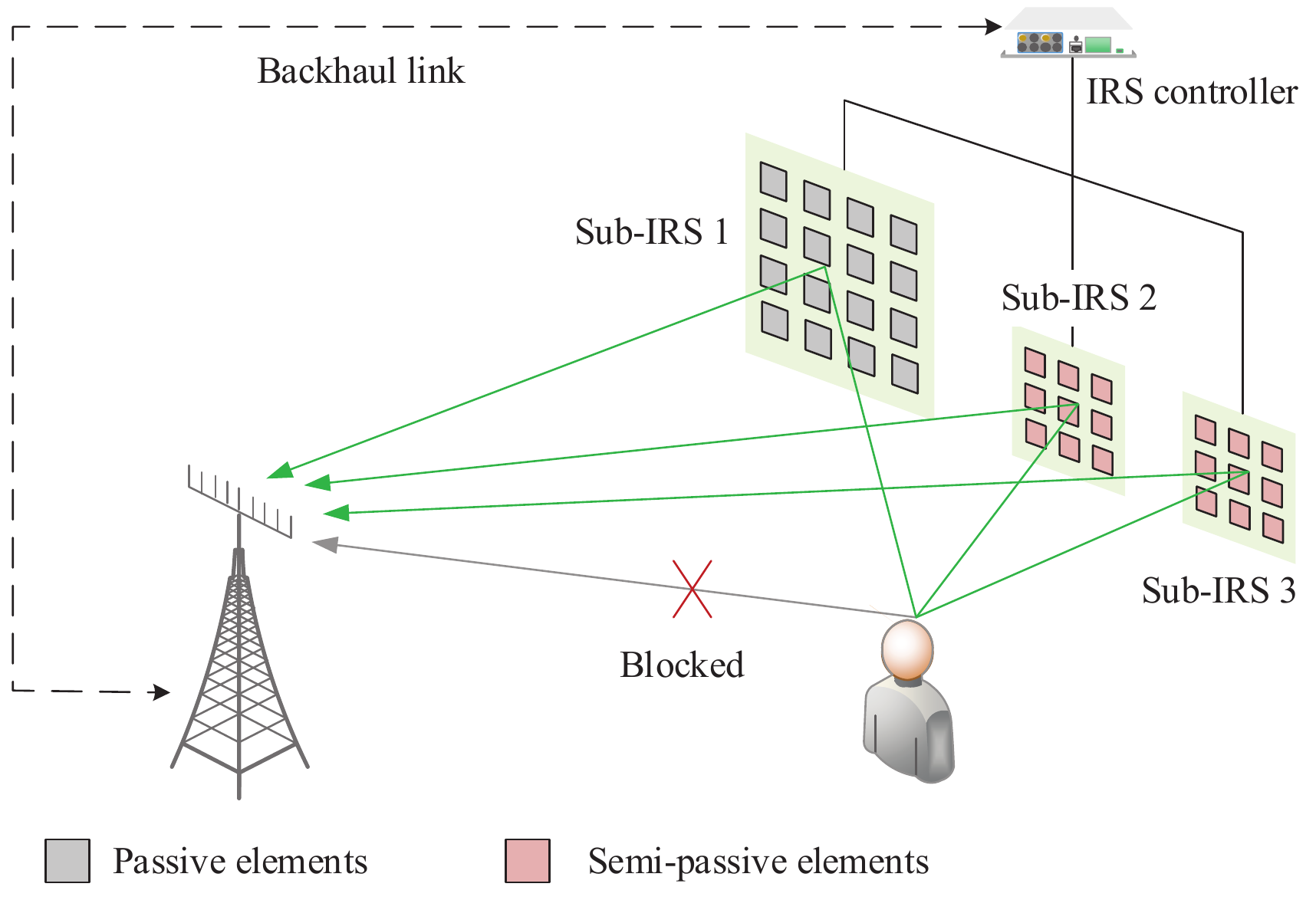}
  \caption{Illustration of  the IRS-based  ISAC system.}
   \label{system_model}
\end{figure}

 As shown in Fig.~\ref{system_model}, we consider an IRS-aided system operating in the mmWave band, where a distributed semi-passive IRS with $M$ reflecting elements assists the uplink transmission between the BS and a single-antenna user. The distributed semi-passive IRS consists of 3 sub-IRSs. The first sub-IRS is passive with $M_1$ passive reflecting elements, while the $i$-th ($i=2,3$)  sub-IRS is semi-passive with $M_i \ll M_1$ semi-passive reflecting elements that are capable of both sensing and reflecting. Herein, the distributed semi-passive IRS architecture is proposed to achieve integrated sensing and communication.
Specifically, the passive sub-IRS assists data transmission by reflecting, and meanwhile the two semi-passive sub-IRSs carry out user positioning by  operating in the sensing mode.
The BS has an $N$-element  uniform linear array (ULA) along the $y$ axis, while the $i$-th sub-IRS has an $M_{y,i}\times M_{z,i}$  uniform rectangular array (URA) lying on the $y$-$o$-$z$ plane.
Moreover, there is a backhaul link for information exchange  between the BS and the IRS. In this paper, the quasi-static block-fading channel is considered for the user-IRS channel, which remains nearly unchanged in each fading block but varies from one block to another. Due to the fixed locations of both the BS and the sub-IRSs, we assume that the channels between the BS and sub-IRSs remain constant over a long period. Furthermore, we assume that the direct link between the BS and the user does not exist due to blockage or unfavorable propagation environment.

\subsection{ISAC Transmission  Protocol}
\begin{figure}[!ht]
  \centering
  \includegraphics[width=6in]{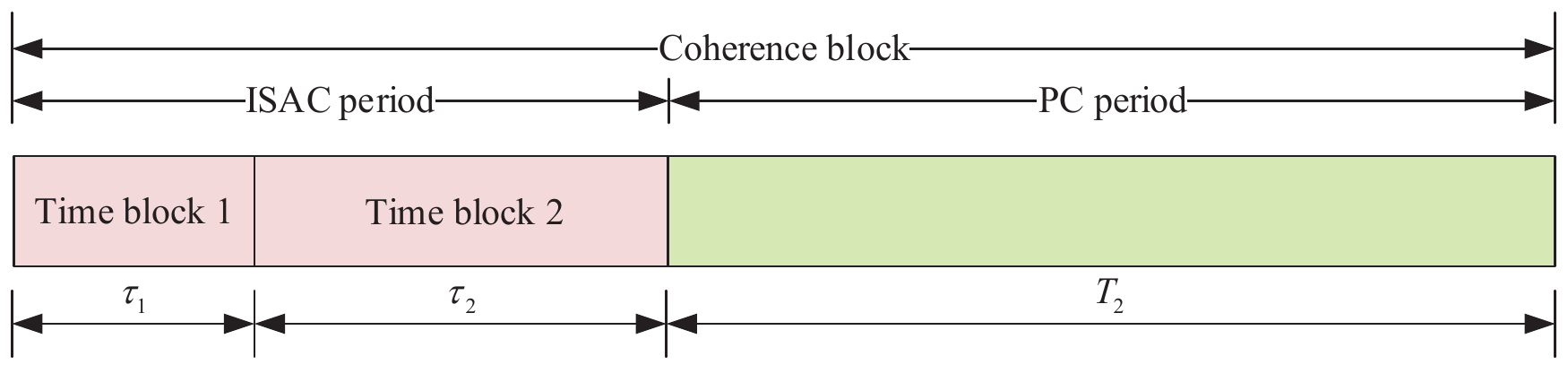}
  \caption{ISAC transmission protocol.}
   \label{transmission_protocol}
\end{figure}

As shown in Fig.\ref{transmission_protocol}, we consider a coherence block, during which the user-IRS channel remains unchanged. Each coherence
block is composed  of two periods, ISAC period with $T_1$ time slots (symbol duration) and PC period with $T_2$ time slots. The ISAC period is divided into two time blocks  with $\tau_1$ and $\tau_2$ time slots, respectively.
During each time block of the ISAC period, the user sends communication signals to the BS. The  passive sub-IRS assists data transmission by reflection, and meanwhile the remaining two semi-passive sub-IRSs operate in the  sensing mode. {\color{black}By using the signals sensed at the two semi-passive sub-IRSs, the IRS carries out the location estimation task and obtains the estimated user's location, which is  then sent to the BS via the backhaul link\footnote{\color{black}In the considered IRS-based ISAC system, sensed location information and IRS phase shifts are exchanged  between the BS and the IRS via the backhaul link. The sensed location information is exchanged once per time block of the ISAC period, while the IRS phase shifts  are exchanged  twice per coherence block.}.
 In the first time block,  the phase shift beam of the passive sub-IRS is randomly generated, due to the unavailability of any CSI knowledge. In the second time block, by using the user's location estimated in the first time block,  the BS carries out the beamforming optimization task and obtains a better phase shift beam, which is then sent to the IRS via the backhaul link.
 During the PC period, the two semi-passive sub-IRSs switch to the reflecting mode to enhance the uplink data transmission. The BS uses more accurate location information acquired  in the second time block of the ISAC period for beamforming optimization and then shares the optimized  phase shifts with the IRS via the backhaul link.}

\begin{remark}
In the first time block, we roughly estimate the user location in a short time so that the communication performance in the second time block can be quickly improved by using the estimated location information for beamforming. In general, the length of the second time block is longer than that of the first time block so that high-accuracy user location can be acquired for more effective beamforming design in the longest PC period.
\end{remark}

\subsubsection{ISAC Period} During the $n$-th time block of the ISAC period, the user sends  $\sqrt{\rho}s(t)$, satisfying {\color{black} $s(t) \sim \mathcal{CN}(0,1)$}, to the BS at  time slot $t\in \mathcal{N}_n=\{(n-1)\tau_1+1 ,\cdots, \tau_1+(n-1)\tau_2 \}$, where $\rho$ is the transmit power. The received signal at the BS via the passive sub-IRS  is
\begin{align} \label{E7}
    { y}(t)=\sqrt{\rho} [ {\bf w}^{(n)}]^H {\bf H}_{\text{I2B},1}{\bm \Theta}_{1}^{(n)} {\bf h}_{\text{U2I},1}s(t) + [ {\bf w}^{(n)}]^H {\bf n}_\text{BS}(t), t\in \mathcal{N}_n,n=1,2,
\end{align}
where ${\bf w}^{(n)}$, satisfying $\| {\bf w}^{(n)} \|=1$, is the BS combining vector in the $n$-th time block, ${\bf H}_{\text{I2B},i} \in \mathbb{C}^{N \times M_i}$ and ${\bf h}_{\text{U2I},i} \in \mathbb{C}^{M_i \times 1} $ are  the channels from the $i$-th sub-IRS to the BS and from the user to the $i$-th sub-IRS, respectively. The phase shift matrix of the first sub-IRS in the $n$-th time block is given by ${\bm \Theta}_{1}^{(n)}=\text{diag}({\bm \xi}_1^{(n)})$, with the phase shift beam being ${\bm \xi}_1^{(n)}={[ e^{j\vartheta_{1,1}^{(n)} },..., e^{j\vartheta_{1,m}^{(n)}},...,  e^{j\vartheta_{1,M_i}^{(n)} }]}^T$. In addition, $\bf{n}_\text{BS} $ is the additive white Gaussian noise (AWGN) at the BS, whose elements follow the complex Gaussian distribution $ \mathcal{CN} \left({ 0},\sigma_0^2  \right)$.


The two semi-passive sub-IRSs operate in the sensing mode, and the received signal at the $i$-th sub-IRS ($i=2,3$)  is given by
\begin{align} \label{E1}
    {\bf x}_i(t)=\sqrt{\rho} {\bf h}_{\text{U2I},i} s(t)+
   \sqrt{\rho} {\bf H}_{\text{I2I},i}{\bm \Theta}_1^{(n)}  {\bf h}_{\text{U2I},1} s(t) +{\bf n}_i(t), i=2,3, t\in \mathcal{N}_n,n=1,2,
\end{align}
where  ${\bf H}_{\text{I2I},i}$ is the channel from the passive sub-IRS (i.e., the first sub-IRS) to the semi-passive sub-IRS (i.e., sub-IRS $i=2,3$). In addition,  ${\bf n}_i$ is the AWGN at the $i$-th sub-IRS, whose elements follow the complex Gaussian distribution $\mathcal{CN} \left({ 0},\sigma_0^2  \right)$.

The instantaneous achievable rate\footnote{\color{black}
In the first time block of the ISAC period, non-cohrent detection is considered, due to the unavailability of any CSI knowledge. The rate achieved by non-coherent detection can be approximated by the rate expression (\ref{Erate}) with perfect CSI \cite{non-coherent}, since the length of the coherence time (in symbols) is much larger than the number of transmit antennas.} during the ISAC period is given by
\begin{align} \label{Erate}
R(t)=\log_2\left( 1+ \frac{ \rho \left| [ {\bf w}^{(n)}]^H {\bf H}_{\text{I2B},1}{\bm \Theta}_{1}^{(n)} {\bf h}_{\text{U2I},1} \right|^2 }{\sigma_0^2}\right) , t\in \mathcal{N}_n,n=1,2.
\end{align}

\subsubsection{PC Period} During the PC period, the user sends  $\sqrt{\rho} s(t)$ to the BS at time slot $t\in \mathcal{T}_2 \triangleq \{T_1+1,\cdots,T_1+T_2\}$. The received signal at the BS via the whole distributed IRS  is
\begin{align} \label{E10}
    y(t)=\sqrt{\rho} {\bf w}^H (t) {\bf H}_{\text{I2B}}{\bm \Theta}(t) {\bf h}_{\text{U2I}}s(t) +{\bf w}^H (t) {\bf n}_\text{BS}(t), t\in \mathcal{T}_2,
\end{align}
where ${\bf H}_{\text{I2B}} \triangleq [{\bf H}_{\text{I2B},1},{\bf H}_{\text{I2B},2},{\bf H}_{\text{I2B},3}] \in \mathbb{C}^{N \times M}$ and ${\bf h}_{\text{U2I}} \triangleq [{\bf h}_{\text{U2I},1}^T,{\bf h}_{\text{U2I},2}^T,{\bf h}_{\text{U2I},3}^T]^T\in \mathbb{C}^{M \times 1} $ are  the channels from the IRS to the BS and from the user to the IRS, respectively. The phase shift matrix is  ${\bm \Theta}=\text{diag}({\bm \xi})$, where the phase shift beam is given by
${\bm \xi}\triangleq [{\bm \xi}_1^T,{\bm \xi}_2^T, {\bm \xi}_3^T]^T$ with the  phase shift beam of the $i$-th sub-IRS being ${\bm \xi}_i={[ e^{j\vartheta_{i,1}},..., e^{j\vartheta_{i,m}},...,  e^{j\vartheta_{i,M_i} }]}^T$.

The instantaneous achievable rate during the PC period is given by
\begin{align}
R(t)=\log_2\left( 1+ \frac{ \rho \left| {\bf w}^H (t) {\bf H}_{\text{I2B}}{\bm \Theta}(t) {\bf h}_{\text{U2I}} \right|^2 }{\sigma_0^2}\right) , t\in \mathcal{T}_2.
\end{align}

\subsection{Channel Model}
In general, the IRS is deployed with line-of-sight (LoS) paths to both the BS and the user. In addition, since the non-line-of-sight (NLoS) path is much weaker than the LoS path for mmWave communications\footnote{For mmWave signals, channel measurement campaigns reveal that  signal power of the LoS component is about 13 dB higher than the sum of power of non-line-of-sight (NLoS) components{\color{black}\cite{muhi2010modelling}}.}, we only consider the dominated LoS paths. Hence, the channels from the $i$-th sub-IRS to the BS is modelled as
\begin{align} \label{E8}
& {\bf H}_{\text{I2B},i}=\alpha_{\text{I2B},i} {\bf a}\left(u_{\text{I2B},i}^\text{A} \right)
{\bf b}_i^H\left(u_{\text{I2B},i}^\text{D},v_{\text{I2B},i}^\text{D} \right), i=1,2,3,
\end{align}
where $\alpha_{\text{I2B},i}$ is the complex channel gain, ${\bf a}$ and ${\bf b}_i$ are array response vectors for the BS and the $i$-th sub-IRS, respectively.
The two effective angles of departure (AoDs) at the $i$-th sub-IRS are defined as
\begin{align}
& u_{\text{I2B},i}^\text{D} =2 \pi \frac{d_\text{IRS} }{\lambda} \cos (\gamma_{\text{I2B},i}^\text{D}) \sin (\varphi_{\text{I2B},i}^\text{D}),\\
& v_{\text{I2B},i}^\text{D}=2 \pi \frac{d_\text{IRS} }{\lambda} \sin (\gamma_{\text{I2B},i}^\text{D}),
\end{align}
where  $d_\text{IRS}$ is the distance between two adjacent reflecting elements, $\lambda$ is the carrier wavelength, $ \gamma_{\text{I2B},i}^\text{D} $ and $\varphi_{\text{I2B},i}^\text{D}$ are the elevation and azimuth AoDs for the link from the $i$-th sub-IRS to the BS, respectively.
The effective angle of arrival (AoA) at the BS is defined as
\begin{align}
u_{\text{I2B},i}^\text{A}=2 \pi \frac{d_\text{BS} }{\lambda} \sin (\theta_{\text{I2B},i}^\text{A}),
\end{align}
where $d_\text{BS}$ is the distance between two adjacent antennas,  and $\theta_{\text{I2B},i}^\text{A}$
is the AoA at the BS.

Similarly, the channels from the user to the $i$-th sub-IRS is modelled as
\begin{align} \label{E2}
    {\bf h}_{\text{U2I},i}= \alpha_{\text{U2I},i} {\bf b}_i \left(u_{\text{U2I},i}^\text{A},v_{\text{U2I},i}^\text{A}\right), i=1,2,3,
\end{align}
where $\alpha_{\text{U2I},i}$ is the complex channel gain, $u_{\text{U2I},i}^\text{A}$ and $v_{\text{U2I},i}^\text{A}$
are two effective angles of arrival (AoAs) from the user to the $i$-th sub-IRS,  which are defined as
\begin{align}
& u_{\text{U2I},i}^\text{A} =2 \pi \frac{d_\text{IRS} }{\lambda} \cos (\gamma_{\text{U2I},i}^\text{A}) \sin (\varphi_{\text{U2I},i}^\text{A}),\\
& v_{\text{U2I},i}^\text{A}=2\pi \frac{d_\text{IRS} }{\lambda} \sin (\gamma_{\text{U2I},i}^\text{A}),
\end{align}
where $ \gamma_{\text{U2I},i}^\text{A} $ and $\varphi_{\text{U2I},i}^\text{A}$ are the elevation and azimuth AoAs for the link from the user to the $i$-th sub-IRS, respectively. Furthermore, we assume that $d_\text{BS}=d_\text{IRS}=\frac{\lambda}{2}$. Then, the array response vectors for the BS and the $i$-th sub-IRS are given by
\begin{align}
   & {\bf a} (u)=\left[1,\cdots, e^{j(n-1)u},\cdots, e^{j(N-1)u} \right]^T, \\
   & {\bf b}_i (u,v)=\left[1,\cdots, e^{j(n-1)u},\cdots, e^{j(M_{y,i}-1)u} \right]^T \otimes \left[1,\cdots, e^{j(m-1)v},\cdots, e^{j(M_{z,i}-1)v} \right]^T.
\end{align}

Also, the channel from the passive sub-IRS to the semi-passive sub-IRS (i.e., sub-IRS $i=2,3$) is modelled as
\begin{align} \label{E3}
{\bf H}_{\text{I2I},i}=\alpha_{\text{I2I},i} {\bf b}_i\left(u_{\text{I2I},i}^\text{A},v_{\text{I2I},i}^\text{A}  \right)
{\bf b}_1^H\left(u_{\text{I2I},i}^\text{D},v_{\text{I2I},i}^\text{D}  \right), i=2,3,
\end{align}
where $\alpha_{\text{I2I},i}$ is the complex channel gain for the link from the first sub-IRS to the $i$-th sub-IRS, and the two  effective AoAs at the $i$-th sub-IRS are defined as
\begin{align}
& u_{\text{I2I},i}^\text{A}=2 \pi \frac{d_\text{IRS} }{\lambda} \cos (\gamma_{\text{I2I},i}^\text{A}) \sin (\varphi_{\text{I2I},i}^\text{A}),\\
& v_{\text{I2I},i}^\text{A}= 2 \pi \frac{d_\text{IRS} }{\lambda} \sin (\gamma_{\text{I2I},i}^\text{A}),
\end{align}
where $ \gamma_{\text{I2I},i}^\text{A} $ and $\varphi_{\text{I2I},i}^\text{A}$ are the elevation and azimuth AoAs from the first sub-IRS to the $i$-th sub-IRS, respectively. The two effective AoDs at the first sub-IRS are defined as
\begin{align}
& u_{\text{I2I},i}^\text{D}=2 \pi \frac{d_\text{IRS} }{\lambda} \cos (\gamma_{\text{I2I},i}^\text{D}) \sin (\varphi_{\text{I2I},i}^\text{D}),\\
& v_{\text{I2I},i}^\text{D}= 2 \pi \frac{d_\text{IRS} }{\lambda} \sin (\gamma_{\text{I2I},i}^\text{D}),
\end{align}
where $ \gamma_{\text{I2I},i}^\text{D} $ and $\varphi_{\text{I2I},i}^\text{D}$ are the elevation and azimuth AoDs from the first sub-IRS to the $i$-th sub-IRS, respectively.


\section{Location Sensing} \label{s2}
During the ISAC period, the two semi-passive sub-IRSs operate in the sensing mode.
By substituting (\ref{E2}) and (\ref{E3}) into (\ref{E1}), the received signal at the $i$-th sub-IRS during the $n$-th time block can be rewritten as
\begin{align}
    {\bf x}_i(t)&=\sqrt{\rho} \alpha_{\text{U2I},i}  {\bf b}_i \left(u_{\text{U2I},i}^\text{A},v_{\text{U2I},i}^\text{A}\right) s(t) \\
    &+ \sqrt{\rho}\alpha_{\text{U2I},1}
    \alpha_{\text{I2I},i} {\bf b}_i\left(u_{\text{I2I},i}^\text{A},v_{\text{I2I},i}^\text{A}  \right)
{\bf b}_1^H\left(u_{\text{I2I},i}^\text{D},v_{\text{I2I},i}^\text{D}  \right)
    {\bm \Theta}_1^{(n)}
   {\bf b}_1 \left(u_{\text{U2I},1}^\text{A},v_{\text{U2I},1}^\text{A}\right)  s(t) +{\bf n}_i(t), t\in \mathcal{N}_n, \nonumber
\end{align}
where the first term is the signal from the user, which involves the user location information, and the second term is the interference from the passive sub-IRS.

The above equation can be expressed in a more compact form
\begin{align}
   {\bf x}_i(t)=\sqrt{\rho}{\bf B}_i {\bm \beta}_i^{(n)} s(t)+{\bf n}_i(t), t\in \mathcal{N}_n,
\end{align}
where
\begin{align}
& {\bf B}_i \triangleq \left[{\bf b}_i \left(u_{\text{U2I},i}^\text{A},v_{\text{U2I},i}^\text{A}\right), {\bf b}_i\left(u_{\text{I2I},i}^\text{A},v_{\text{I2I},i}^\text{A}  \right) \right] \in \mathbb{C}^{M_i \times 2}, \\
&{\bm \beta}_i^{(n)}\triangleq \left[\alpha_{\text{U2I},i},  \
\alpha_{\text{U2I},1}
    \alpha_{\text{I2I},i}{\bf b}_1^H\left(u_{\text{I2I},i}^\text{D},v_{\text{I2I},i}^\text{D}  \right)
    {\bm \Theta}_1^{(n)}
   {\bf b}_1 \left(u_{\text{U2I},1}^\text{A},v_{\text{U2I},1}^\text{A}\right) \right]^T  \in \mathbb{C}^{2 \times 1} .
\end{align}


Next, according to a sequence of $  {\bf x}_i(t), t \in \mathcal{N}_n$, we try to  estimate the two pairs of effective AoAs for links from the user to  the $i$-th sub-IRS and from the passive sub-IRS to the $i$-th sub-IRS, and then distinguish  $\left(u_{\text{U2I},i}^\text{A},v_{\text{U2I},i}^\text{A}\right)$ from these two pairs of effective AoAs.
Finally, according to the effective AoAs from the user to the two semi-passive sub-IRSs (i.e., $\left(u_{\text{U2I},i}^\text{A},v_{\text{U2I},i}^\text{A}\right),i=2,3$), the location of the user is determined.

\subsection{Estimate Effective AoAs} \label{III.A}
By using  total least square (TLS) estimation of signal parameters via rotational
invariance technique (ESPRIT) method, we separately estimate the effective AOAs corresponding to the $y$ axis (i.e., $u_{\text{U2I},i}^\text{A}$ and $u_{\text{I2I},i}^\text{A}$) and the $z$ axis (i.e., $v_{\text{U2I},i}^\text{A}$ and $v_{\text{I2I},i}^\text{A}$). Then, invoking the multiple signal classification (MUSIC) method, we pair the effective AOAs corresponding to the $y$ axis with those corresponding to the $z$ axis.

\begin{figure}[!ht]
  \centering
  \includegraphics[width=4.5in]{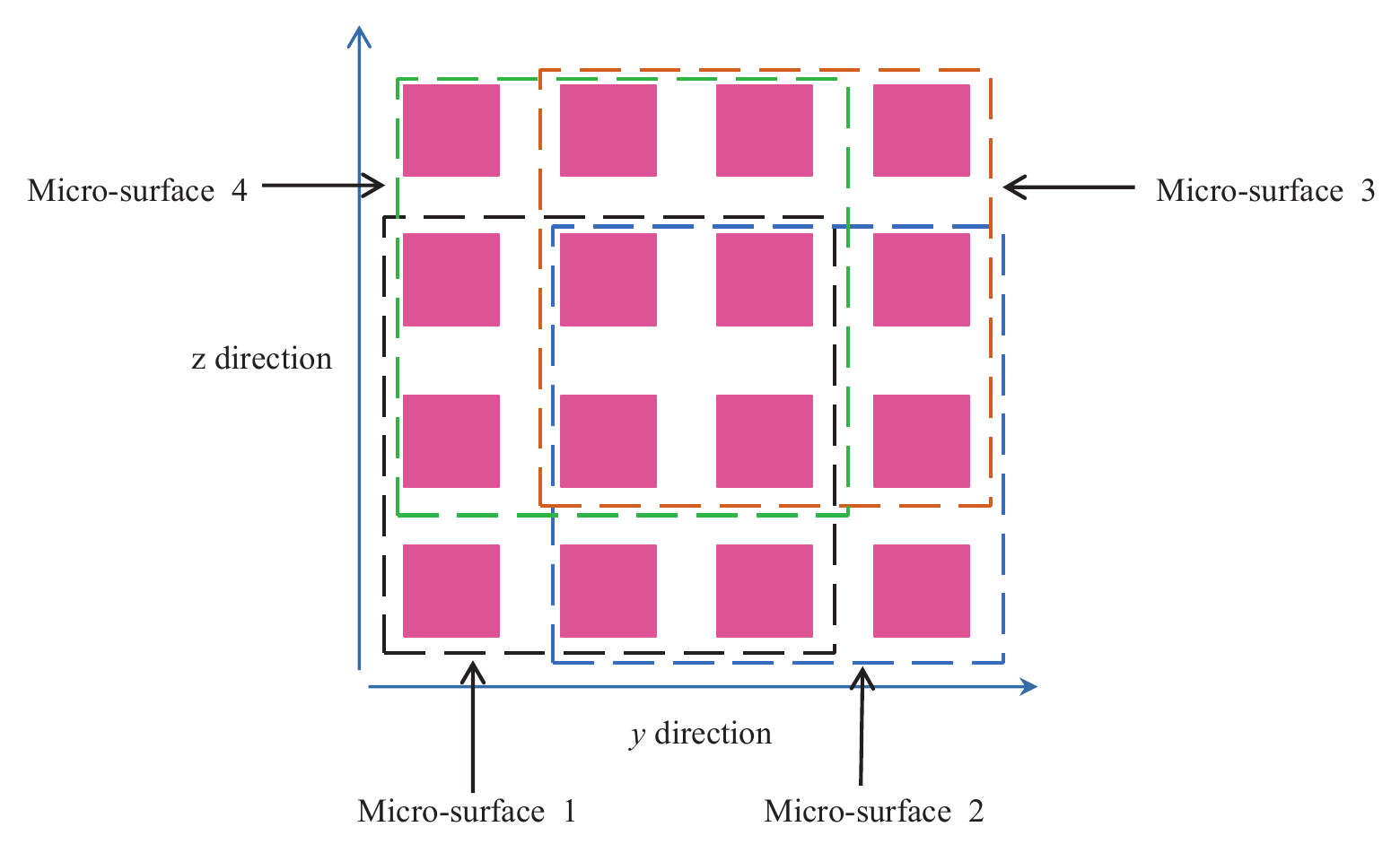}
  \caption{An example  of the micro-surface.}
  \label{subsurface}
\end{figure}

Without loss of generality, we focus on the  estimation of the effective AoAs at the $i$-th sub-IRS in the $n$-th time block ($i \in \{2,3\}$ and $n \in \{1,2\}$).  To remove the coherency of the received signals, we use  forward-backward spatial smoothing (FBSS){\color{black} \cite{van2004optimum}} to preprocess the signals received at the  semi-passive sub-IRS.
Specifically, for the $i$-th sub-IRS, we construct a set of $N_{\text{micro},i}$ micro-surfaces each with $L_{\text{micro},i}= Q_{y,i} \times Q_{z,i}$ semi-passive elements.  Each micro-surface is shifted by one row along the $z$ direction or one  column along the $y$ direction from the preceding micro-surface. An example is shown in Fig.\ref{subsurface}, where we construct a set of 4 micro-surfaces  each with $3 \times 3$ semi-passive elements.

The received signals at the $m$-th micro-surface of the $i$-th sub-IRS during the $n$-th time block are denoted by  $ {\bf x}_{i,m} (t) \in {{\mathbb{C}}^{L_{\text{micro},i} \times 1}}, t \in \mathcal{N}_n$.
Following the FBSS technique, the auto-correlation matrix of ${\bf x}_{i,m} (t), t \in \mathcal{N}_n$, i.e.,
${\bf R}_{i}^{(n)} \triangleq \mathbb{E} \{{\bf x}_{i,m} (t)  [{\bf x}_{i,m} (t)]^H \}$, can be estimated as
\begin{align}
 \hat{\bf R}_{i}^{(n)} =\frac{1}{2 \tau_n N_{\text{micro},i} }
  \sum\limits_{t \in \mathcal{N}_n } \sum\limits_{m=1}^{N_{\text{micro},i}}
  \left\{ {\bf x}_{i,m} (t)  [{\bf x}_{i,m} (t)]^H
   + {\bf J} [ {\bf x}_{i,m} (t) ]^* [ {\bf x}_{i,m}(t) ]^T {\bf J} \right\},
\end{align}
where ${\bf J}$ is the exchange matrix, with the 1 elements residing on its counterdiagonal and all other elements being zero.

Then, perform
eigenvalue decomposition of ${{\hat{\bf R}}}_{i}^{(n)}$
\begin{align}
 \hat{\bf R}_{i}^{(n)}= {\bf U}_i^{(n)}\text{diag}\left(\lambda_{i,1}^{(n)},\dots,\lambda_{i,L_{ \text{micro},i }}^{(n)}\right) [{\bf U}_i^{(n)}]^H,
\end{align}
where $ {\bf U}_i^{(n)} \triangleq [{\bf u}_{i,1}^{(n)}, \dots, {\bf u}_{i,L_{\text{micro},i}}^{(n)}]$ and the eigenvalues $\lambda_{i,1}^{(n)},\dots,\lambda_{i,L_{\text{micro},i}}^{(n)}$ are in a descending order.

\subsubsection{ Estimate $u_{\text{U2I},i}^\text{A}$ and $u_{\text{I2I},i}^\text{A}$ by using the TLS ESPRIT algorithm} 
\begin{figure}[htbp]
\begin{minipage}[t]{1 \linewidth}
\centering
\subfigure[]{ \label{subsurface_uy}
\includegraphics[width=4in]{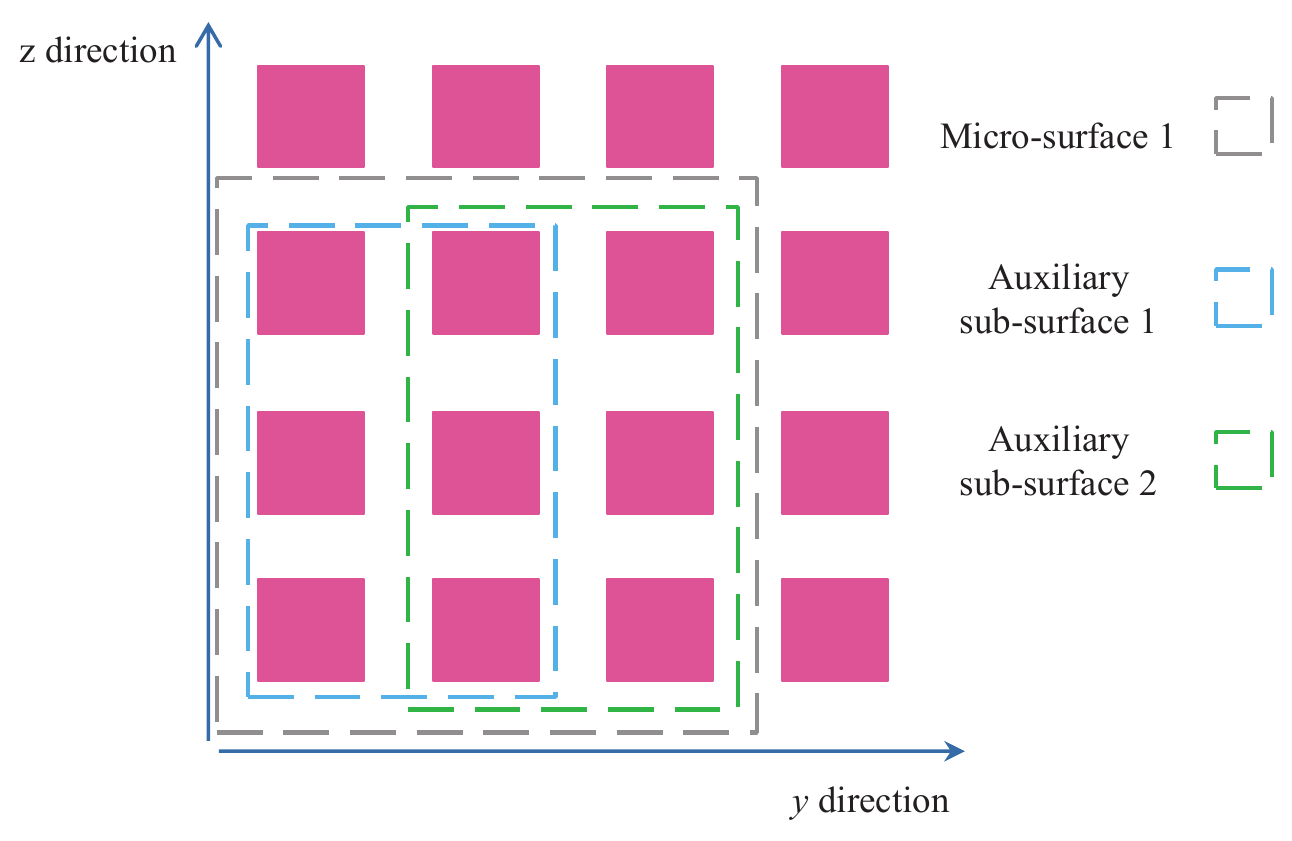}}
\end{minipage}
\begin{minipage}[t]{1 \linewidth}
\centering
\subfigure[]{ \label{subsurface_uz}
\includegraphics[width=4in]{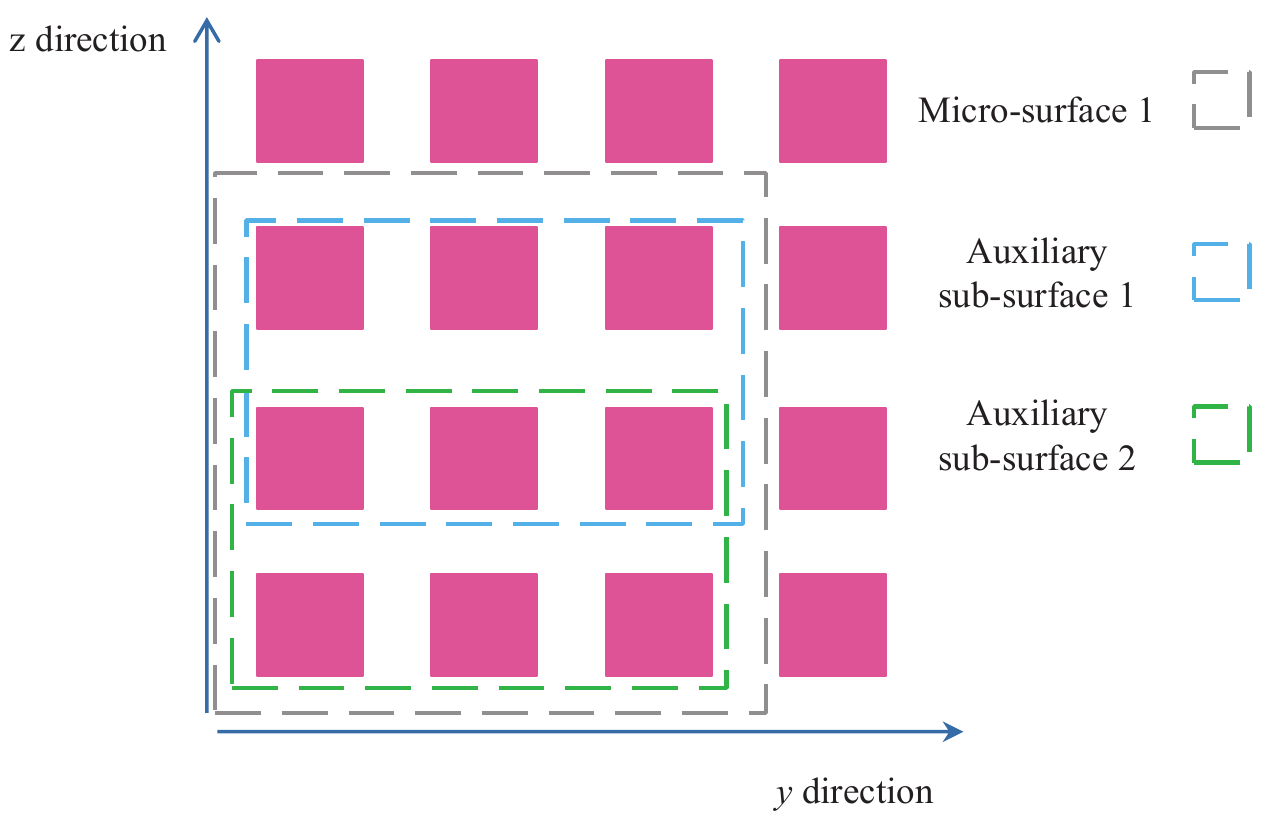}}
\end{minipage}
 \caption{An example of the auxiliary sub-surface.}
\label{simulation1}
\end{figure}

For the $i$-th sub-IRS,  we construct two
auxiliary sub-surfaces of its first micro-surface, with the size of $L_{\text{aux},i}=(Q_{y,i}-1)\times Q_{z,i}$, as illustrated in Fig. ~\ref{subsurface_uy}.
The signal sub-space corresponding to the two auxiliary sub-surfaces is\footnote{\color{black}There are two pairs of AOAs seen at each semi-pasive sub-IRS, one corresponding to the link from user to the semi-passive sub-IRS, and another corresponding to the link from the passive sub-IRS to the semi-passive sub-IRS. Hence, the matrix ${\bf U}_{\text{S},ik}^{(n)}$ has two columns.}
\begin{align}
 {\bf U}_{\text{S},ik}^{(n)} \triangleq   {\bf J}_{k} {\bf U}_{\text{S},i}^{(n)}, k=1,2,
 \end{align}
where  ${\bf U}_{\text{S},i}^{(n)} \triangleq [{\bf u}_{i,1}^{(n)}, {\bf u}_{i,2}^{(n)} ] \in \mathbb{C}^{L_{\text{micro},i} \times 2} $ and  ${\bf J}_{k} \in \mathbb{R}^{L_{\text{aux},i} \times L_{\text{micro},i}}$ ($k\in\{1,2\}$) is a selecting matrix, whose elements are either 1 or 0. If the $j$-th reflecting element of the micro-surface 1 is selected as the $i$-th element of the auxiliary sub-surface $k\in\{1,2\}$, then we set $[{\bf J}_{ k}]_{ij}=1$. Otherwise, we set $[{\bf J}_{k}]_{ij}=0$.

Calculate
\begin{align}
    {\bm \Phi}_{\text{TLS},i}^{(n)}=-{\bf V}_{i,12}^{(n)} [{\bf V}_{i,22}^{(n)}]^{-1},
\end{align}
where ${\bf V}_{i,12}^{(n)}$ and ${\bf V}_{i,22}^{(n)}$ are $2 \times 2$ matrices defined by the  eigendecomposition of the $4 \times 4$ matrix
\begin{align}
  & {\bf C}_i^{(n)} \triangleq \left[ {\bf U}_{\text{S},i1}^{(n)},  {\bf U}_{\text{S},i2}^{(n)} \right]^H \left[ {\bf U}_{\text{S},i1}^{(n)},  {\bf U}_{\text{S},i2}^{(n)}\right]
   \\
   &=
   \begin{bmatrix}
{\bf V}_{i,11}^{(n)} & {\bf V}_{i,12}^{(n)}\\
{\bf V}_{i,21}^{(n)} & {\bf V}_{i,22}^{(n)}
\end{bmatrix}
{\bm \Lambda}_{C,i}^{(n)}\begin{bmatrix}
{\bf V}_{i,11}^{(n)} & {\bf V}_{i,12}^{(n)}\\
{\bf V}_{i,21}^{(n)} & {\bf V}_{i,22}^{(n)}
\end{bmatrix}^H, \nonumber
\end{align}
where ${\bm \Lambda}_{C,i}^{(n)} \triangleq \text{diag}\left( \lambda_{\text{C},i1}^{(n)},\dots, \lambda_{\text{C},i4}^{(n)}\right) $ with the eigenvalues in a decreasing order.

Perform eigendecomposition of $ {\bm \Phi}_{\text{TLS},i}^{(n)}$ and obtain its eigenvalues $\lambda_{\text{TLS},il}^{(n)},l=1,2$.
Then, we have the two effective AoAs at the $i$-th sub-IRS corresponding to the $y$ axis estimated as
\begin{align}
    \check{u}_{il}^{(n)}=\text{angle}(\lambda_{\text{TLS},il}^{(n)}),l=1,2.
\end{align}
It is worth noting that the estimators of  $u_{\text{U2I},i}^\text{A}$ and  $u_{\text{I2I},i}^\text{A}$ in the $n$-th time block belong to $\mathcal{U}_i^{(n)}\triangleq \{\check{u}_{il}^{(n)},l=1,2 \}$, i.e., $\hat{u}_{\text{U2I},i}^{\text{A},(n)},\hat{u}_{\text{I2I},i}^{\text{A},(n)} \in \mathcal{U}_i^{(n)} $.

\subsubsection{Estimate $v_{\text{U2I},i}^\text{A}$ and $v_{\text{I2I},i}^\text{A}$ by using TLS ESPRIT algorithm}  For the $i$-th sub-IRS,  we construct two
auxiliary sub-surfaces of its first micro-surface, with the size of $\tilde{L}_{\text{aux},i}=Q_{y,i} \times (Q_{z,i}-1)$, as illustrated in Fig.~\ref{subsurface_uz}. Following the similar process of estimating $u_{\text{U2I},i}^\text{A}$ and $u_{\text{I2I},i}^\text{A}$, we estimate the two effective AoAs at the $i$-th sub-IRS corresponding to the $z$ axis  as $\check{v}_{il}^{(n)},l=1,2$. And the estimators of  $v_{\text{U2I},i}^\text{A}$ and  $v_{\text{I2I},i}^\text{A}$ in the $n$-th time block belong to $\mathcal{V}_i^{(n)} \triangleq \{\check{v}_{il}^{(n)},l=1,2 \}$, i.e., $\hat{v}_{\text{U2I},i}^{\text{A},(n)},\hat{v}_{\text{I2I},i}^{\text{A},(n)} \in \mathcal{V}_i^{(n)} $.

\subsubsection{Pair $\check{u}_{il}^{(n)}$ and  $\check{v}_{il}^{(n)}$ by using the MUSIC algorithm}
Let
\begin{align}
 \check{f}_{i,ls}^{(n)}\triangleq {\bf b}_{ \text{micro},i }^H (  \check{u}_{il}^{(n)}, \check{v}_{is}^{(n)})
  {\bf U}_{\text{N},i}^{(n)},
\end{align}
 where ${\bf U}_{\text{N},i}^{(n)} \triangleq [{\bf u}_{i,3}^{(n)}, \dots, {\bf u}_{i, L_{\text{micro},i} }^{(n)} ] \in \mathbb{C}^{L_{\text{micro},i} \times (L_{\text{micro},i}-2)} $, and ${\bf b}_{ \text{micro},i }$ is the array response vector of the micro-surface on the $i$-th sub-IRS.  Then compute
\begin{align}
  f\left( \check{u}_{il}^{(n)}, \check{v}_{is}^{(n)}\right)=
  \check{f}_{i,ls}^{(n)} [\check{f}_{i,ls}^{(n)}]^H, l,s=1,2,
\end{align}
 and choose the two minima
 $f\left( \hat{u}_{il}^{(n)}, \hat{v}_{il}^{(n)}\right), l=1,2$, where $\hat{u}_{il}^{(n)} \in \mathcal{U}_i^{(n)}$ and  $\hat{v}_{il}^{(n)} \in \mathcal{V}_i^{(n)}$. As such, we obtain two pairs of effective AoAs $\left( \hat{u}_{il}^{(n)}, \hat{v}_{il}^{(n)}\right), l=1,2$.

\subsubsection{Determine $\left(u_{\text{U2I},i}^\text{A},v_{\text{U2I},i}^\text{A} \right)$}
Since there are two pairs of effective AoAs, we need to determine which pair is corresponding to  $\left(u_{\text{U2I},i}^\text{A},v_{\text{U2I},i}^\text{A} \right)$.

Denote the locations of the BS and  the $i$-th sub-IRS by   ${\bf q}_\text{BS}=(x_\text{BS},y_\text{BS},z_\text{BS})$ and  ${\bf q}_i=(x_i,y_i,z_i)$, respectively.
Since the locations of the BS and all the sub-IRSs are fixed, we assume that these locations are perfectly known.
By invoking the available location information, the effective AoAs from the first sub-IRS to the $i$-th sub-IRS can be calculated as
\begin{align}
   & u_{\text{I2I},i}^\text{A}= \frac{y_i-y_1}{\| {\bf q}_i-{\bf q}_1\|},\\
  & v_{\text{I2I},i}^\text{A}= \frac{z_i-z_1}{\| {\bf q}_i-{\bf q}_1\|}.
\end{align}

Finally, excluding the effective AoA pair  corresponding to $\left(u_{\text{I2I},i}^\text{A},v_{\text{I2I},i}^\text{A} \right) $ from  $\left\{ \left( \hat{u}_{il}^{(n)},\hat{v}_{il}^{(n)}\right)|  l=1,2\right\}$,  the remaining one is the estimator for  $\left(u_{\text{U2I},i}^{\text{A},(n)},v_{\text{U2I},i}^{\text{A},(n)} \right) $, i.e., $\left(\hat{u}_{\text{U2I},i}^{\text{A},(n)},\hat{v}_{\text{U2I},i}^{\text{A},(n)} \right)$.

\subsection{Estimate User Location} \label{III.B}
In the following, we try to estimate the user location according to the estimated effective AoAs
$\left(u_{\text{U2I},i}^{\text{A},(n)},v_{\text{U2I},i}^{\text{A},(n)} \right),i=2,3$. The locations of the sub-IRSs and the user have the following relationship with these effective AoAs:
\begin{align}
 & \hat{u}_{\text{U2I},i}^{\text{A},(n)}=\frac{ y_i-\hat{y}_\text{U}^{(n)}  }{\hat{d}_{\text{U2I},i}^{(n)} }, i=2,3,\\
 & \hat{v}_{\text{U2I},i}^{\text{A},(n)}=\frac{ z_i-\hat{z}_\text{U}^{(n)} }{\hat{d}_{\text{U2I},i}^{(n)} }, i=2,3,
\end{align}
where $\hat{\bf q}_\text{U}^{(n)}=[\hat{x}_\text{U}^{(n)}, \hat{y}_\text{U}^{(n)},\hat{z}_\text{U}^{(n)}]^T$ denotes the  user location estimated in the $n$-th time block, and $\hat{d}_{\text{U2I},i}^{(n)} \triangleq \|\hat{\bf q}_\text{U}^{(n)}-{\bf q}_i\|$ is the estimated distance between the user and the $i$-th sub-IRS.

The above equations can be expressed in a more compact form
\begin{align}
    {\bf A}^{(n)}\hat{\bf z}^{(n)}={\bf p},
\end{align}
where
\begin{align}
& {\bf A}^{(n)} \triangleq
\begin{pmatrix}
1 & 0 & \hat{u}_{\text{U2I},2}^{\text{A},(n)} &  0\\
0 & 1 & \hat{v}_{\text{U2I},2}^{\text{A},(n)} &  0\\
1 & 0 & 0 &  \hat{u}_{\text{U2I},3}^{\text{A},(n)}\\
0 & 1 & 0 &  \hat{v}_{\text{U2I},3}^{\text{A},(n)}
\end{pmatrix},\\
   & {\bf z}^{(n)} \triangleq
   \left[ \hat{y}_\text{U}^{(n)}, \hat{z}_\text{U}^{(n)}, \hat{d}_{\text{U2I},2}^{(n)},\hat{d}_{\text{U2I},3}^{(n)}\right]^T, \\
   & {\bf p} \triangleq \left[y_2,z_2,y_3,z_3 \right]^T.
\end{align}

By solving the above matrix equation, we obtain
\begin{align}
  &\hat{d}_{\text{U2I},2}^{(n)}=
  \frac{\hat{u}_{\text{U2I},3}^{\text{A},(n)} (z_2-z_3) -\hat{v}_{\text{U2I},3}^{\text{A},(n)} (y_2-y_3)  }
  {\hat{u}_{\text{U2I},3}^{\text{A},(n)} \hat{v}_{\text{U2I},2}^{\text{A},(n)} - \hat{u}_{\text{U2I},2}^{\text{A},(n)} \hat{v}_{\text{U2I},3}^{\text{A},(n)}    },  \\
  & \hat{d}_{\text{U2I},3}^{(n)}=
  \frac{\hat{u}_{\text{U2I},2}^{\text{A},(n)} (z_3-z_2) -\hat{v}_{\text{U2I},2}^{\text{A},(n)} (y_3-y_2)  }
  {\hat{u}_{\text{U2I},2}^{\text{A},(n)} \hat{v}_{\text{U2I},3}^{\text{A},(n)} - \hat{u}_{\text{U2I},3}^{\text{A},(n)} \hat{v}_{\text{U2I},2}^{\text{A},(n)}    },\\
  & \hat{y}_\text{U}^{(n)}=y_2-\hat{u}_{\text{U2I},2}^{\text{A},(n)} \hat{d}_{\text{U2I},2}^{(n)}, \label{E4}\\
  & \hat{z}_\text{U}^{(n)}=z_2-\hat{v}_{\text{U2I},2}^{\text{A},(n)} \hat{d}_{\text{U2I},2}^{(n)}\label{E5}.
\end{align}

Then, we calculate  $\hat{x}_\text{U}^{(n)}$. Noticing that $\hat{x}_\text{U}^{(n)}$ satisfies
\begin{align}
  \left(\hat{x}_\text{U}^{(n)}-x_i \right)^2 =\left(\hat{d}_{\text{U2I},i}^{(n)}\right)^2-\left(\hat{y}_\text{U}^{(n)}-y_i \right)^2-\left(\hat{z}_\text{U}^{(n)}-z_i \right)^2,i=2,3,
\end{align}
we have
\begin{align}
 \hat{x}_\text{U}^{(n)}=   \underset {\omega_2 }{\text{arg}} \ \  \underset{\omega_2 \in \{x_2 \pm d_{x,2}\}, \ \omega_3 \in \{x_3 \pm d_{x,3}\} }{ \text{min} } \left|\omega_2-\omega_3 \right|, \label{E6}
\end{align}
where
\begin{align}
     d_{x,i} \triangleq \sqrt{ \left(\hat{d}_{\text{U2I},i}^{(n)}\right)^2-\left(\hat{y}_\text{U}^{(n)}-y_i \right)^2-\left(\hat{z}_\text{U}^{(n)}-z_i \right)^2}.
\end{align}

Combining (\ref{E4}), (\ref{E5}) and (\ref{E6}), we obtain the user location $\hat{\bf q}_\text{U}^{(n)}=[\hat{x}_\text{U}^{(n)}, \hat{y}_\text{U}^{(n)},\hat{z}_\text{U}^{(n)}]^T$.

{\color{black}
\begin{remark}
To estimate the  user location, some IRS elements are equipped with receiving radio frequency (RF) chains, which incur additional power consumption and hardware cost. However, since the proportion of IRS elements with  receiving RF chains is very small, the resulting increase in power consumption and hardware cost will not be much.
\end{remark}
}

\section{Beamforming Design}\label{s3}
In this section, the estimated user location is used for beamforming design in both the ISAC period and the PC period.

\subsection{ISAC period}
In the ISAC period, only the passive  sub-IRS (i.e., the first sub-IRS) operates in the reflecting mode to assist uplink data transmission.

By substituting (\ref{E8}) and (\ref{E2}) into (\ref{E7}), the received signal at the BS can be rewritten as
\begin{align}
 y(t)=& \sqrt{\rho} \alpha_{\text{I2B},1} \alpha_{\text{U2I},1}  [ {\bf w}^{(n)}]^H  {\bf a}\left(u_{\text{I2B},1}^\text{A} \right)
{\bf b}_1^H\left(u_{\text{I2B},1}^\text{D},v_{\text{I2B},1}^\text{D} \right){\bm \Theta}_{1}^{(n)}  {\bf b}_1 \left(u_{\text{U2I},1}^\text{A},v_{\text{U2I},1}^\text{A}\right) s(t)\\
&+[ {\bf w}^{(n)}]^H{\bf n}_\text{BS}(t), t\in \mathcal{N}_n,n=1,2. \nonumber
\end{align}

We aim to maximize the received signal power at the BS, and the optimization problem is formulated as
\begin{subequations}
\begin{align}
    &\underset{{\bf w}^{(n)},{\bm \xi}_1^{(n)}}{\text{max}} && \left| [ {\bf w}^{(n)}]^H  {\bf a}\left(u_{\text{I2B},1}^\text{A} \right)
{\bf b}_1^H\left(u_{\text{I2B},1}^\text{D},v_{\text{I2B},1}^\text{D} \right){\bm \Theta}_{1}^{(n)}  {\bf b}_1 \left(u_{\text{U2I},1}^\text{A},v_{\text{U2I},1}^\text{A}\right) \right|^2,\\
&\text{s.t.}  &&\left\| {\bf w}^{(n)} \right\|=1,\\
& && |[{\bm \xi}_1^{(n)}]_i|=1,i=1,\cdots,M_1,
\end{align}
\end{subequations}
which can be equivalently decomposed into  two sub-problems, i.e., the sub-problem corresponding to the BS combining vector
\begin{subequations}
\begin{align}
   &\underset{{\bf w}^{(n)}}{\text{max}} \ \  \left| [ {\bf w}^{(n)}]^H  {\bf a}\left(u_{\text{I2B},1}^\text{A} \right) \right|^2,\\
&\text{s.t.}  \ \ \ \  \left\| {\bf w}^{(n)} \right\| =1,
\end{align}
\end{subequations}
and the sub-problem corresponding to the phase shift beam of the first sub-IRS
\begin{subequations}
\begin{align}
    &\underset{{\bm \xi}_1^{(n)}}{\text{max}} \ \  \left |
{\bf b}_1^H\left(u_{\text{I2B},1}^\text{D},v_{\text{I2B},1}^\text{D} \right){\bm \Theta}_{1}^{(n)}  {\bf b}_1 \left(u_{\text{U2I},1}^\text{A},v_{\text{U2I},1}^\text{A}\right) \right |^2,\\
&\text{s.t.}   \ \ \ \ |[{\bm \xi}_1^{(n)}]_i|=1,i=1,\cdots,M_1.
\end{align}
\end{subequations}

It can be easily verified that the optimal solutions for the above sub-problems are
\begin{align}
 & {\bf w}^{(n)}=  \frac{1}{\sqrt{N}} {\bf a}\left(u_{\text{I2B},1}^\text{A} \right), \label{E9}\\
 & {\bm \xi}_1^{(n)}=\text{diag}\left( {\bf b}_1^* \left(u_{\text{U2I},1}^\text{A},v_{\text{U2I},1}^\text{A}\right) \right)
 {\bf b}_1\left(u_{\text{I2B},1}^\text{D},v_{\text{I2B},1}^\text{D} \right).\label{E10}
\end{align}

According to the locations of the BS and the first IRS, $u_{\text{I2B},1}^\text{A}$ in (\ref{E9}) is calculated as
\begin{align}
  u_{\text{I2B},1}^\text{A}=\frac{ y_\text{BS}-y_1 }{ \left\|{\bf q}_\text{BS}-{\bf q}_1 \right\|}.
\end{align}

Due to  unavailability of  user location information during the first time block, the phase shift beam of the first sub-IRS, i.e., ${\bm \xi}_1^{(1)}$, is randomly selected in the first time block.
Hence, in the following, we only focus on the phase shift design of the first sub-IRS in the second time block (i.e., ${\bm \xi}_1^{(2)}$) by invoking the user location estimated in the first time block. As such, we have
\begin{align}
   {\bm \xi}_1^{(2)}=\text{diag}\left( {\bf b}_1^* \left(\hat{u}_{\text{U2I},1}^{\text{A},(1)},\hat{v}_{\text{U2I},1}^{\text{A},(1)}\right) \right)
 {\bf b}_1\left({u}_{\text{I2B},1}^{\text{D}},{v}_{\text{I2B},1}^{\text{D}} \right),
\end{align}
where the effective AoAs and AoDs  are estimated according to the location of the first sub-IRS and the  user location estimated in the first time block, which are given by
\begin{align}
 & \hat{u}_{\text{U2I},1}^{\text{A},(1)}=\frac{ \hat{y}_{\text{U}}^{(1)}-y_1  }{ \left\|\hat{\bf q}_{\text{U}}^{(1)}-{\bf q}_1  \right\|}, \\
  & \hat{v}_{\text{U2I},1}^{\text{A},(1)}=\frac{ \hat{z}_{\text{U}}^{(1)}-z_1  }{ \left\|\hat{\bf q}_{\text{U}}^{(1)}-{\bf q}_1  \right\|}, \\
  &  {u}_{\text{I2B},1}^{\text{D}}=-\frac{ {y}_{\text{BS}}-y_1  }{ \left\|{\bf q}_{\text{BS}}-{\bf q}_1  \right\|}, \\
   & {v}_{\text{I2B},1}^{\text{D}}=-\frac{ {z}_{\text{BS}}-z_1  }{ \left\|{\bf q}_{\text{BS}}-{\bf q}_1  \right\|}.
\end{align}

\subsection{PC Period}
In the PC period, all the three sub-IRSs operate in the reflecting mode to assist the uplink data transmission between the BS and the user. Their phase shifts are designed according to the user location estimated in the second time block of the ISAC period.
By substituting (\ref{E8}) and (\ref{E2}) into (\ref{E10}), the received signal at the BS via the whole distributed IRS  is
\begin{align} \label{E11}
    y(t)=&\sum\limits_{i=1}^{3} \sqrt{\rho} \alpha_{\text{I2B},i} \alpha_{\text{U2I},i}  {\bf w}^H  {\bf a}\left(u_{\text{I2B},i}^\text{A} \right)
{\bf b}_i^H\left(u_{\text{I2B},i}^\text{D},v_{\text{I2B},i}^\text{D} \right){\bm \Theta}_{i}(t)  {\bf b}_i \left(u_{\text{U2I},i}^\text{A},v_{\text{U2I},i}^\text{A}\right) s(t)\\
&+\widetilde{n}_\text{BS}(t), t\in \mathcal{T}_2, \nonumber
\end{align}
where we define $\widetilde{n}_\text{BS}(t) \triangleq {\bf w}^H(t) {\bf n}_\text{BS}(t) $.

Since the distances among the three sub-IRSs are much smaller than the distances between them and the BS, we have the following approximation
$u_{\text{I2B},i}^\text{A} \approx  u_{\text{I2B},1}^\text{A}, i=2,3$. In addition, the number of reflecting elements of the first sub-IRS is much larger than that of the $i$-th sub-IRS ($i=2,3$). Hence, the BS combining vector is designed to combing signals from the direction of the first sub-IRS, i.e.,  $
    {\bf w}=\frac{1}{N}{\bf a}\left(u_{\text{I2B},1}^\text{A} \right)
$.

As such, (\ref{E11}) becomes
\begin{align}
    y(t)=&\sum\limits_{i=1}^{3} \zeta_i
{\bf b}_i^H\left(u_{\text{I2B},i}^\text{D},v_{\text{I2B},i}^\text{D} \right){\bm \Theta}_{i} (t)  {\bf b}_i \left(u_{\text{U2I},i}^\text{A},v_{\text{U2I},i}^\text{A}\right) s(t)+\widetilde{n}_\text{BS}(t), t\in \mathcal{T}_2,
\end{align}
where $\zeta_i \triangleq \sqrt{\rho} \alpha_{\text{I2B},i} \alpha_{\text{U2I},i}  {\bf a}^H\left(u_{\text{I2B},1}^\text{A} \right)  {\bf a}\left(u_{\text{I2B},i}^\text{A} \right)$.

\begin{figure}[!ht]
  \centering
  \includegraphics[width=4.5in]{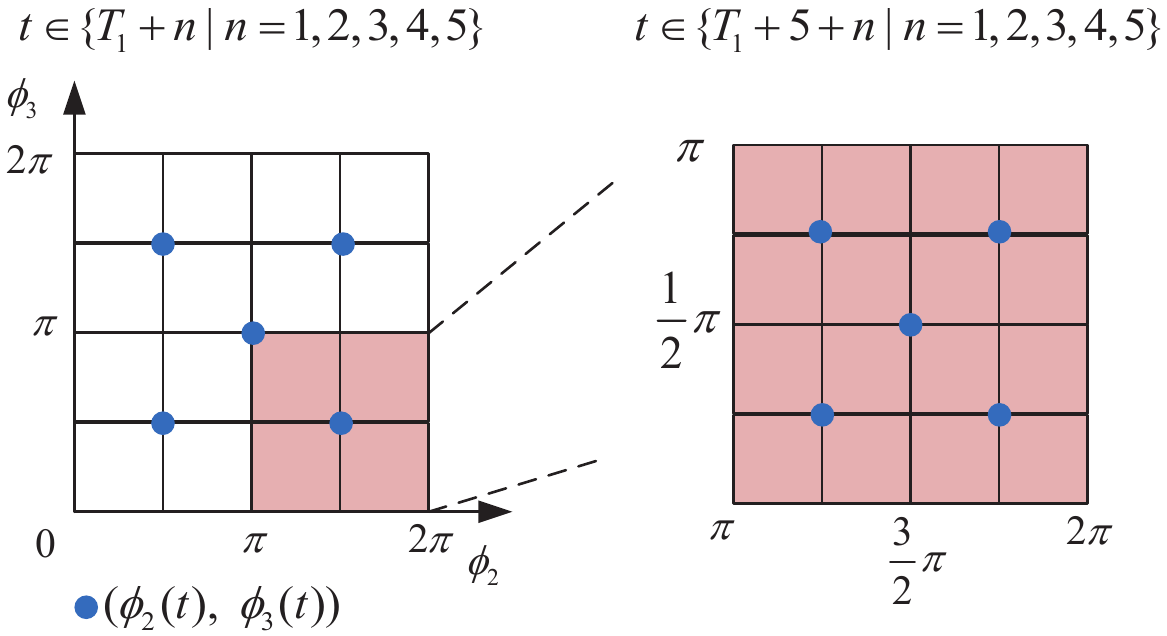}
  \caption{Illustration of $\left( \phi_2(t), \phi_3(t)\right)$ update.}
   \label{phase_update}
\end{figure}

{\color{black}To maximize the received signal power at the BS, we formulate the following optimization problem:
\begin{subequations} \label{E12}
\begin{align}
    &\underset{{\bm \xi}_i(t)}{\text{max}}  \ \  \left| \sum\limits_{i=1}^{3} \zeta_i
{\bf b}_i^H\left(u_{\text{I2B},i}^\text{D},v_{\text{I2B},i}^\text{D} \right){\bm \Theta}_{i}(t)  {\bf b}_i \left(u_{\text{U2I},i}^\text{A},v_{\text{U2I},i}^\text{A}\right) \right|^2,\\
&\text{s.t.} \ \ \ \
 | [{\bm \xi}_i(t)]_s|=1,s=1,\cdots,M_i, i=1,2,3. \label{E14} 
\end{align}
\end{subequations}

Although being non-convex, the above problem admits a closed-form solution by exploiting the special structure of its objective function. Specifically, the objective function satisfies the following inequality
\begin{align} \label{inequality}
  \left|  \sum\limits_{i\!=\!1}^{3} \zeta_i
{\bf b}_i^H (u_{\text{I2B},i}^\text{D},v_{\text{I2B},i}^\text{D} ){\bm \Theta}_{i} (t)  {\bf b}_i (u_{\text{U2I},i}^\text{A},v_{\text{U2I},i}^\text{A}) \right|
\! \le \!  \sum\limits_{i\!=\!1}^{3} \left| \zeta_i
{\bf b}_i^H(u_{\text{I2B},i}^\text{D},v_{\text{I2B},i}^\text{D} ){\bm \Theta}_{i}(t)  {\bf b}_i (u_{\text{U2I},i}^\text{A},v_{\text{U2I},i}^\text{A}) \right|,
\end{align}
where the equality holds if and only if
\begin{align} \label{E13}
 &\text{angle}\left\{  \zeta_i
{\bf b}_i^H\left(u_{\text{I2B},i}^\text{D},v_{\text{I2B},i}^\text{D} \right){\bm \Theta}_{i}(t)  {\bf b}_i \left(u_{\text{U2I},i}^\text{A},v_{\text{U2I},i}^\text{A}\right)
  \right\}
\\
&=
 \text{angle}\left\{ \zeta_i
{\bf b}_j^H\left(u_{\text{I2B},j}^\text{D},v_{\text{I2B},j}^\text{D} \right){\bm \Theta}_{j} (t) {\bf b}_j \left(u_{\text{U2I},j}^\text{A},v_{\text{U2I},j}^\text{A}\right)
\right\}, i\ne j, i,j\in\{1,2,3\}. \nonumber
\end{align}

Next, we show that there always exists a solution ${\bm \xi}(t)\triangleq [ {\bm \xi}_1^T(t),{\bm \xi}_2^T(t),{\bm \xi}_3^T(t)]^T$ that satisfies (\ref{inequality}) with equality as well as the phase shift constraint (\ref{E14}).
With (\ref{inequality}), the problem (\ref{E12}) can be transformed into
\begin{subequations}
\begin{align}
    &\underset{{\bm \xi}_i(t)}{\text{max}}  \ \  \sum\limits_{i=1}^{3} \left| \zeta_i
{\bf b}_i^H\left(u_{\text{I2B},i}^\text{D},v_{\text{I2B},i}^\text{D} \right){\bm \Theta}_{i} (t)  {\bf b}_i \left(u_{\text{U2I},i}^\text{A},v_{\text{U2I},i}^\text{A}\right) \right|^2,\\
&\text{s.t.} \ \ \ \ (\ref{E13}),
 |[{\bm \xi}_i(t)]_s|=1,s=1,\cdots,M_i,i=1,2,3.
\end{align}
\end{subequations}

It is easily verified that the optimal ${\bm \xi}_i(t)$ is
\begin{align}
    {\bm \xi}_i(t)= \text{diag}\left\{ {\bf b}_i^H \left(u_{\text{U2I},i}^\text{A},v_{\text{U2I},i}^\text{A}\right)  \right\}  {\bf b}_i\left(u_{\text{I2B},i}^\text{D},v_{\text{I2B},i}^\text{D} \right) e^{j \phi_i(t)
    }, i\in\{1,2,3\},
\end{align}
where $\phi_1(t)=0$,  $\phi_2(t)=\text{angle}(\zeta_2)-\text{angle}(\zeta_1)$ and $\phi_3(t)=\text{angle}(\zeta_3)-\text{angle}(\zeta_1)$.

Using the location information estimated in the second time block of the ISAC period,  we design ${\bm \xi}_i(t)$ ($t \in \mathcal{T}_2$) as
\begin{align} \label{E15}
    {\bm \xi}_i(t)= \text{diag}\left\{ {\bf b}_i^H \left(\hat{u}_{\text{U2I},i}^{\text{A},(2)},\hat{v}_{\text{U2I},i}^{\text{A},(2)}\right)  \right\}  {\bf b}_i\left(u_{\text{I2B},i}^\text{D},v_{\text{I2B},i}^\text{D} \right) e^{j \phi_i(t)
    }.
\end{align}

However, since $\zeta_{i}$  is unknown, $\left( \phi_2(t), \phi_3(t)\right)$ cannot be determined directly. 
Hence, in the following, we propose a low-complexity bisection-based phase shift beam training algorithm to find the optimal phase tuple $\left( \phi_2(t), \phi_3(t)\right)$ that maximizes the received power at the BS.

As shown in Fig.\ref{phase_update}, during the first $5$ time slots of the PC period, the IRS generates  phase shift beams according to (\ref{E15}) with the phase tuples $\left( \phi_2(t), \phi_3(t)\right), t\in \{T_1+n|n=1,\cdots,5\}$ set to be $\left( \pi, \pi\right)$, $\left( \frac{1}{2}\pi, \frac{1}{2}\pi\right)$, $\left( \frac{1}{2}\pi, \frac{3}{2}\pi\right)$, $\left( \frac{3}{2}\pi, \frac{1}{2}\pi\right)$ and $\left( \frac{3}{2}\pi, \frac{3}{2}\pi\right)$, respectively.
The BS feeds back the phase tuple with the largest BS receiving power to the IRS every five time slots, based on which the IRS updates its phase tuples $\left( \phi_2(t), \phi_3(t)\right)$ in the next 5 time slots and generates the corresponding  phase shift beams according to (\ref{E15}). Specifically, if the phase tuple $\left( \phi_2(t^\star_k), \phi_3(t^\star_k)\right)$ achieves the largest BS receiving power among the five phase tuples $ \left( \phi_2( t), \phi_3( t)\right),{t} \in \{T_1+5(k-1)+n|n=1,\cdots,5\}$, the phase tuples in the next 5 time slots 
$\left( \phi_2(t), \phi_3(t)\right), t\in \{T_1+5k+n| n=1,\cdots,5\}$  are updated as 
\begin{align}
&\left( \phi_2(t^\star_{k}), \phi_3(t^\star_{k}) \right),\\
&\left( \phi_2(t^\star_{k})\pm \frac{\pi}{2^{k-1}}, \phi_3(t^\star_{k}) \pm \frac{\pi}{2^{k-1}} \right).
\end{align} 
For example, as shown in  Fig. \ref{phase_update}, we assume the phase tuple $\left( \frac{3}{2}\pi, \frac{1}{2}\pi\right)$ achieves the largest BS receiving power. Then,  the phase tuples in the next $5$ time slots are updated as  $\left( \frac{3}{2}\pi, \frac{1}{2}\pi\right)$, $\left( \frac{5}{4}\pi, \frac{1}{4}\pi\right)$, $\left( \frac{5}{4}\pi, \frac{3}{4}\pi\right)$, $\left( \frac{7}{4}\pi, \frac{1}{4}\pi\right)$ and $\left( \frac{7}{4}\pi, \frac{3}{4}\pi\right)$.
The detailed  process is given by  Algorithm \ref{alg1}.
}

\begin{algorithm}
\caption{\color{black} Bisection-Based Phase Shift Beam Training Algorithm}
\label{alg1}
{\color{black}
\begin{algorithmic}
\State {\bf Initialization:} Counting index $k=1$, tolerance $\epsilon>0$, and the phase tuples during the first 5 time slots of the PC period  $\left( \phi_2(t), \phi_3(t)\right), t\in \{T_1+n | n=1,\cdots,5\}$ are set as   $\left( \pi, \pi\right)$, $\left( \frac{1}{2}\pi, \frac{1}{2}\pi\right)$, $\left( \frac{1}{2}\pi, \frac{3}{2}\pi\right)$, $\left( \frac{3}{2}\pi, \frac{1}{2}\pi\right)$ and $\left( \frac{3}{2}\pi, \frac{3}{2}\pi\right)$.
\Do
\State The IRS generates  phase shift beams  according to (\ref{E15}) with the phase tuples $\left( \phi_2(t), \phi_3(t)\right), t\in \{T_1+5(k-1)+n | n=1,\cdots,5\}$. 

\State Obtain the received signal power at the BS as $\| {\bf y}\left( \phi_2( t), \phi_3( t)\right) \|^2,{t} \in \{T_1+5(k-1)+n|n=1,\cdots,5\}$.

\State  Find the phase tuple  with the largest received power
\begin{align}
\left( \phi_2(t^\star_k), \phi_3(t^\star_k)\right) =\text{arg} \underset{ \left( \phi_2(t), \phi_3(t)\right),{t} \in \{T_1+5(k-1)+n|n=1,\cdots,5\} }{ \text{max} } \| { \bf y}\left( \phi_2( t), \phi_3( t)\right) \|^2.
\end{align}
\State Let $k\longleftarrow k+1$.
\State Update the phase tuples in the next 5 time slots  $\left( \phi_2(t), \phi_3(t)\right), t\in \{T_1+5(k-1)+n| n=1,\cdots,5\}$  as
\begin{align}
&\left( \phi_2(t^\star_{k-1}), \phi_3(t^\star_{k-1}) \right),\\
&\left( \phi_2(t^\star_{k-1})\pm \frac{\pi}{2^{k}}, \phi_3(t^\star_{k-1}) \pm \frac{\pi}{2^{k}} \right).
\end{align}
\doWhile{ $\| {\bf y}\left( \phi_2(t^\star_{k-1}), \phi_3(t^\star_{k-1}) \right)\|^2-\|{\bf y}\left( \phi_2(t^\star_{k-2}), \phi_3(t^\star_{k-2}) \right) \|^2 > \epsilon$ }

\State In the remaining time slots  $t\in \{T_1+5(k-1), T_1+5(k-1)+1,...,T_1+T_2\}$, design the phase shift beams as (\ref{E15}) with $\left( \phi_2(t), \phi_3(t)  \right)=\left( \phi_2(t^\star_{k-1}), \phi_3(t^\star_{k-1})  \right)$ .
\end{algorithmic}
}
\end{algorithm}

    


{\color{black}
\begin{remark}
 The computational complexity of the proposed beamforming algorithm is mainly determined by the calculation of  (\ref{E15}). As such, the complexity is $\mathcal{O}\left(\sum\limits_{i=1}^{3}M_i^2 \right) $.
\end{remark}
 }
{\color{black}
\section{Extension of the proposed IRS-based ISAC framework to more general cases}
\subsection{General Channel Model}
The proposed location sensing scheme with the LOS channel model can be easily extended to the general channel model with both LOS and NLOS paths. By applying the proposed AOA estimation scheme (see Section \ref{III.A}), the AOA pairs corresponding to both LOS and NLOS paths between the user and each semi-passive sub-IRS can be estimated. Then, according to these estimated AOA pairs, their corresponding path losses can also be obtained. Noticing that the LOS path has the smallest path loss, we can distinguish the AOA pair corresponding to the LOS path, based on which the user location can be determined by invoking the results provided in Section \ref{III.B}. In addition, by ignoring the weak  NLOS components, our proposed beamforming algorithms based on the LOS channel model can be directly applied to the case with a general mmWave channel model. Of course, this would cause a performance loss. 

\subsection{Multi-User Case}
Although the proposed IRS-based ISAC framework only considers the single-user case, it can be extended to the multi-user case by scheduling multiple users on different time-frequency blocks. For each user, communication and location sensing are conducted, sharing  the same time and frequency resources.
As such, the proposed location sensing scheme and  beamforming algorithms for the single-user case can be directly applied to the multi-user case.
}

\section{Simulation Results} \label{s4}
\begin{figure}[!ht]
  \centering
  \includegraphics[width=3.5in]{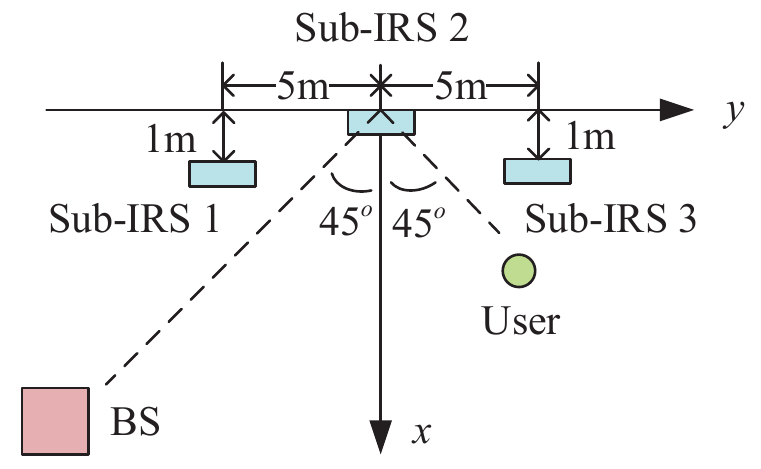}
  \caption{Simulation setup (top view).}
  \label{simulation_model}
\end{figure}

In this section, we provide simulation results to demonstrate the effectiveness of the proposed ISAC transmission protocol. The simulation setup is shown in Fig. \ref{simulation_model}, where the user is on the horizontal floor,
the BS is $20$ meter (m) above the horizontal floor, and the three sub-IRSs are respectively $5$~m, $7$~m and $8$~m above the horizontal floor.
The distances from the BS to the second sub-IRS and from the second sub-IRS to the user are set to be $d_{\text{B2I},2}=50$ m and $d_{\text{I2U},2}=6$ m, respectively.
The  path loss exponents from the IRS to the BS, from the user to the IRS, and from one sub-IRS to other sub-IRSs are set as  $2.3$, $2.2$ and $2.1$ respectively. The path loss at the reference distance of 1 m is set as 30 dB.
Unless otherwise specified, the following setup is used: $N=8$,  $M_1=16 \times 16$, $M_2=M_3=M_\text{semi}=4 \times 4$,  $T=1200$, $T_1=120$, $\tau_1=20$, $\rho=20$dBm and  noise power $\sigma_0^2=-80$ dBm.

\subsection{Performance of Location Sensing}
Without loss of generality, we focus on the performance of  location sensing during the first time block of the ISAC period. { \color{black} To this end, we adopt the root mean square error (RMSE) of the estimated user's location as the performance metric, given by
$\varepsilon=\sqrt{ \mathbb{E}\{  \|  \hat{\bf q}_{\text{U}}^{(1)}-{\bf q}_{\text{U}}\|^2}\}$
}.

\begin{figure}[!ht]
  \centering
  \includegraphics[width=4.5in]{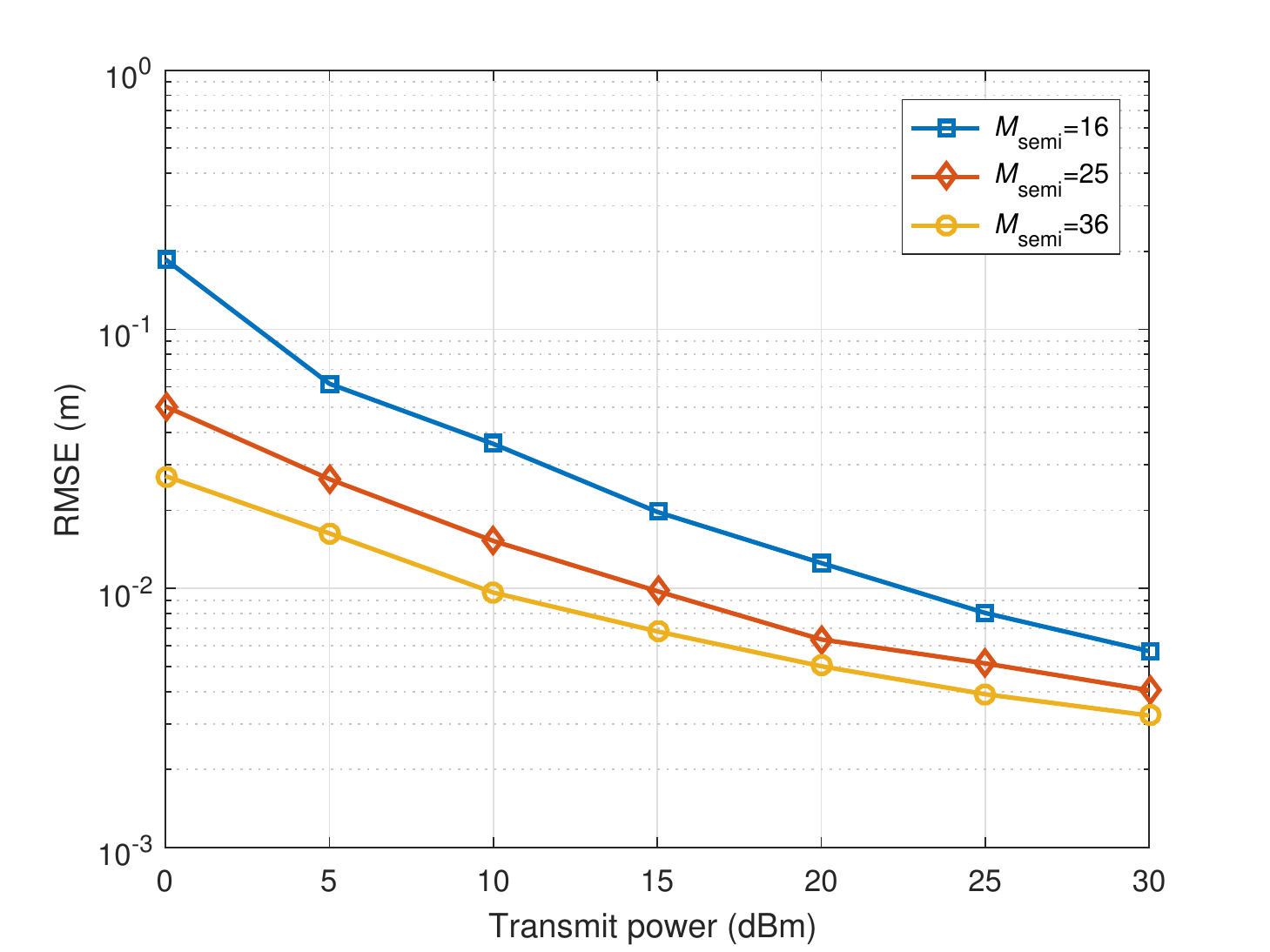}
  \caption{\color{black}RMSE of the estimated user location.}
  \label{main1}
\end{figure}

Fig.\ref{main1}  shows the RMSE of the estimated user location under different transmit power. As can be readily seen, the RMSE of the estimated user location decreases with the transmit power. Moreover, increasing the number of semi-passive elements can significantly reduce the  required transmit power. For instance, for the positioning accuracy of $10^{-2}$ m, the required transmit power drops from about $ 22$ dBm to $10 $ dBm, when the number of semi-passive elements (per semi-passive sub-IRS) increases from 16 to 36.

\begin{figure}[!ht]
  \centering
  \includegraphics[width=4.5in]{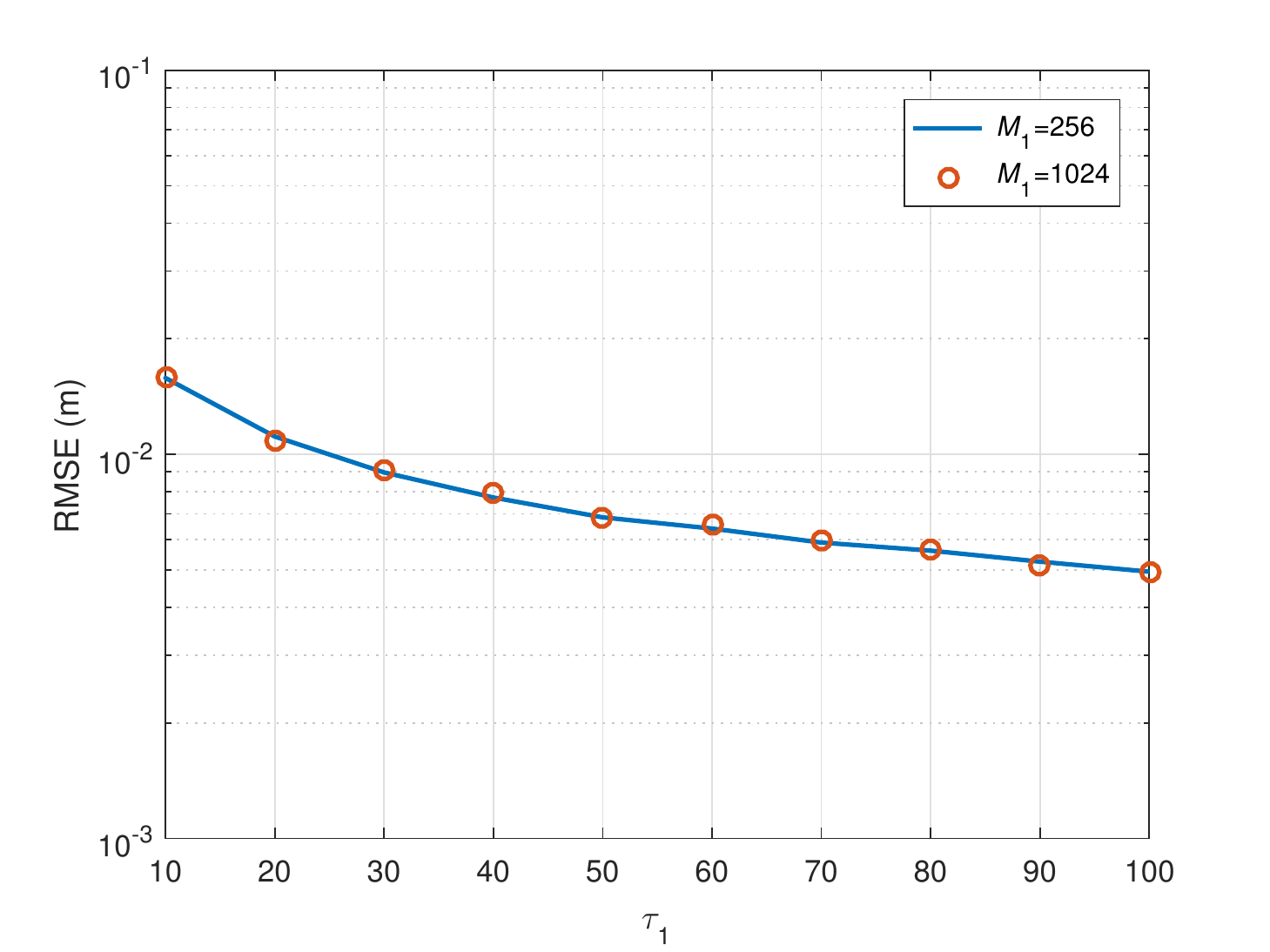}
  \caption{\color{black}Impact of sensing time on the location sensing accuracy.}
  \label{main2}
\end{figure}

Fig.\ref{main2} shows the impact of the length of sensing time (i.e., $\tau_1$) on the RMSE. Since collecting more data helps suppress the adverse effect of noise,  the accuracy of location sensing improves with the length of sensing time. In addition, although increasing the number of passive reflecting elements will cause more interference to the semi-passive sub-IRSs, which perform location sensing during the ISAC period, the accuracy of location sensing will not be affected. This is because the proposed location sensing method can effectively eliminate the interference from the passive sub-IRS.

\begin{figure}[!ht]
  \centering
  \includegraphics[width=4.5in]{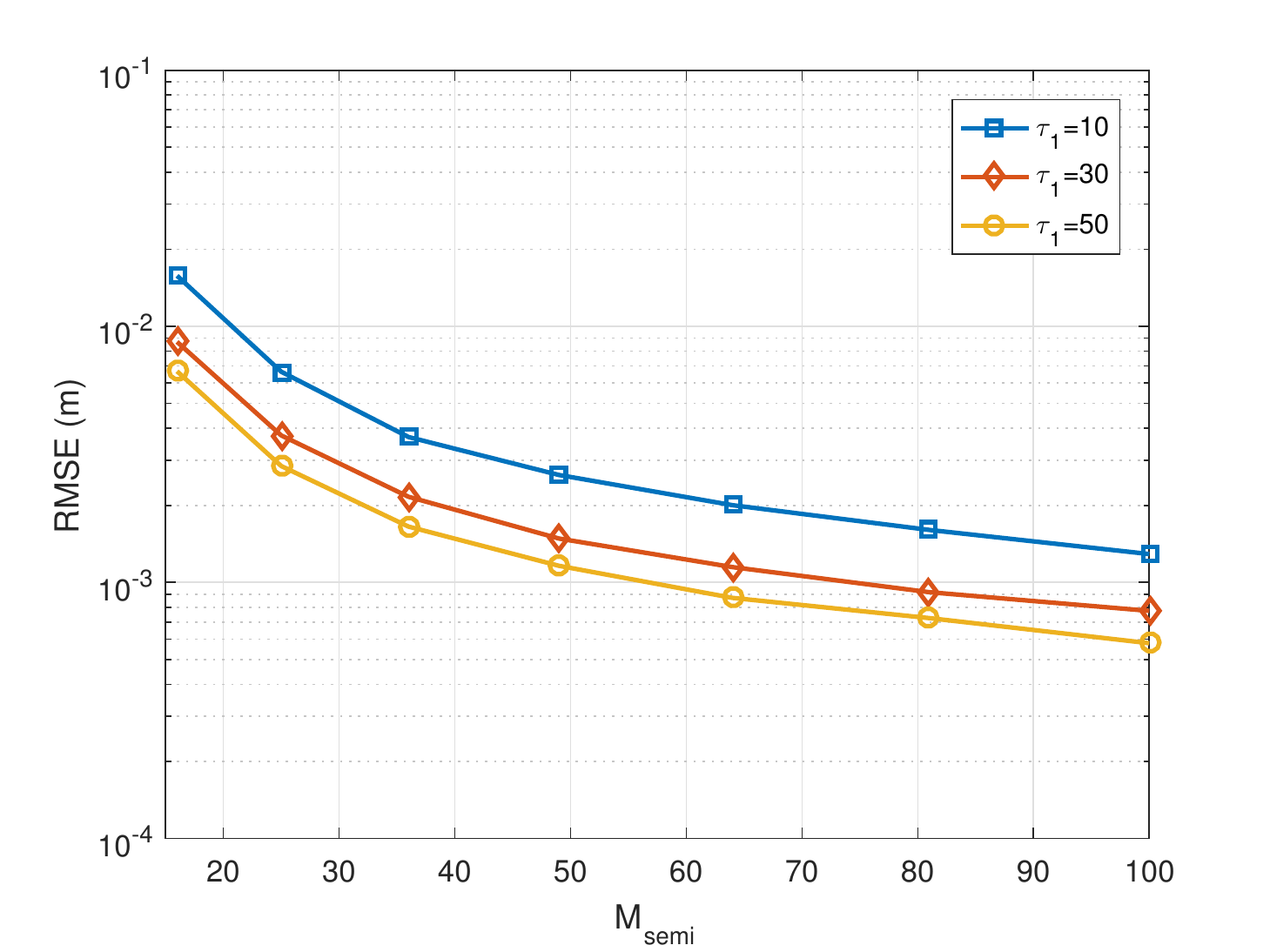}
  \caption{\color{black}Impact of the number of  semi-passive elements on the location sensing accuracy.}
  \label{main3}
\end{figure}

Fig.\ref{main3} shows the impact of the number of  semi-passive elements (per semi-passive sub-IRS) on the accuracy of location sensing. As can be seen, the positioning accuracy improves as  the number of  semi-passive elements (per semi-passive sub-IRS) increases. In addition,  a high positioning accuracy can be achieved even with a small number of semi-passive elements. For instance, for a given sensing time $\tau_1=30$, a  millimeter-level position accuracy is achieved with only $25$ semi-passive elements (per semi-passive sub-IRS).
Moreover, increasing the number of semi-passive elements can  shorten the time of location sensing. For example, for a give positioning accuracy of $5$ mm, the length of sensing time (i.e., $\tau_1$) drops from $30$  to $10$, when the number of  semi-passive elements (per semi-passive sub-IRS) increases from  $16$ to $25$.

\begin{figure}[!ht]
  \centering
  \includegraphics[width=4.5in]{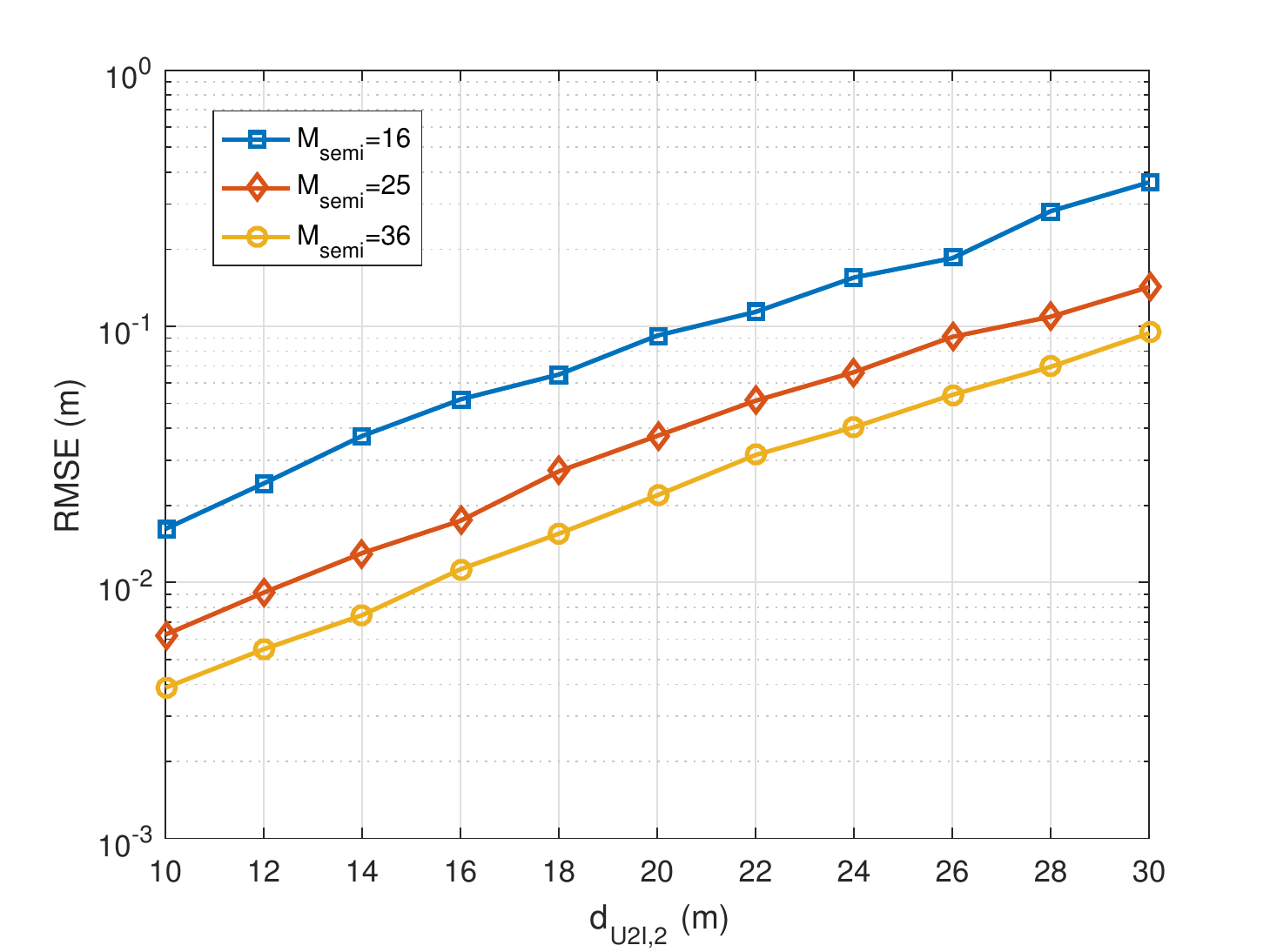}
  \caption{\color{black}Impact of the IRS-user distance on the location sensing accuracy.}
  \label{main4}
\end{figure}

Fig.\ref{main4} presents the RMSEs under different distances between the IRS and the user.  The performance of location sensing degrades with the increased distance from the user to the IRS, since the signals received by  two semi-passive sub-IRSs become weaker. This performance degradation can be compensated by adding more semi-passive elements. By increasing the number of semi-passive element (per semi-passive sub-IRS) from $25$ to $36$, the RMSE remains unchanged at $ 10^{-2}$ m when the distance between the user and the IRS  increases  from $12$ m to  $16$ m.

\subsection{Performance of the Proposed Beamforming Algorithms}
In this subsection, we will provide numerical results to verify the effectiveness of the proposed beamforming  algorithms in the ISAC and PC periods, respectively.

\begin{figure}[!ht]
  \centering
  \includegraphics[width=4.5in]{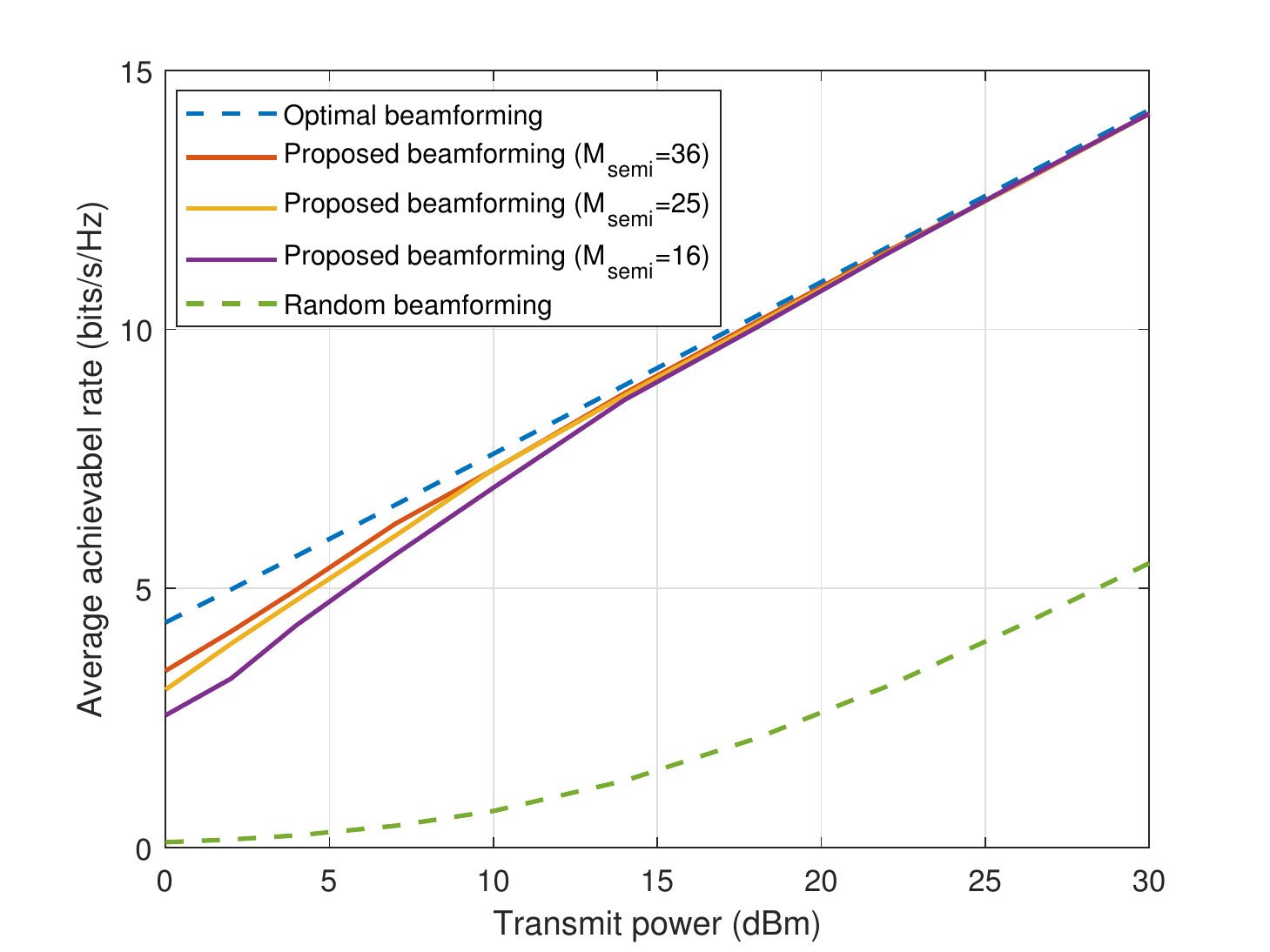}
  \caption{Performance of the proposed beamforming scheme in the ISAC period.}
  \label{main5}
\end{figure}

Fig.\ref{main5} presents the performance of the proposed beamforming scheme in the ISAC period, where the average achievable rate is defined as
{\color{black}
\begin{align}
    \bar{R}_\text{ISAC}=\frac{1}{T_1} \sum\limits_{t=1}^{T_1} R (t).
\end{align}
}
The optimal scheme with perfect CSI and  the random scheme with randomly generated  phase shifts are presented as two benchmarks. The proposed beamforming scheme performs much better than the random scheme, and achieves  similar performance to the optimal scheme. {\color{black}This is because the optimal beamforming is determined by location information, which can be accurately estimated by using the proposed location sensing scheme.} Moreover, as the transmit power increases, the gap between the proposed scheme and  the optimal scheme gradually vanishes.  Also, increasing the number of semi-passive elements  improves the performance of the proposed scheme, due to more accurate location sensing.

\begin{figure}[!ht]
  \centering
  \includegraphics[width=4.5in]{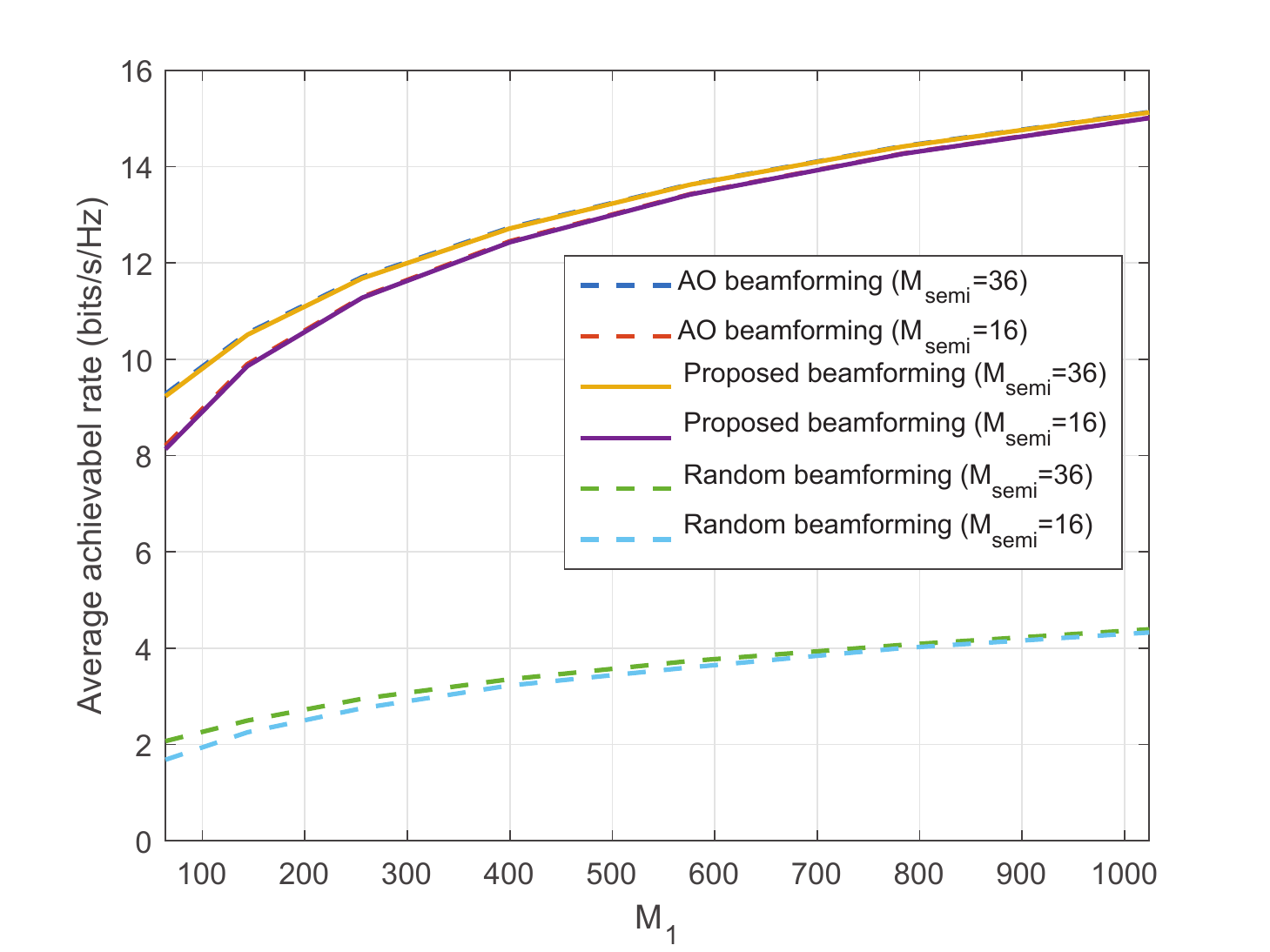}
  \caption{Performance of the proposed beamforming scheme in the PC period.}
  \label{main6}
\end{figure}

Fig.\ref{main6} presents the performance of the proposed beamforming scheme in the PC period, where the average achievable rate is defined as
{\color{black}
\begin{align}
    \bar{R}_\text{PC}=\frac{1}{T_2} \sum\limits_{t=T_1+1}^{T_1+T_2} R (t).
\end{align}
}
For comparison, the performance of the AO scheme with perfect CSI \cite{qingqing1} and  the random scheme with  randomly generated phase shifts are presented.
The proposed beamforming scheme is superior to the random scheme, and achieves almost the same performance as the AO beamforming scheme with perfect CSI.  All the three beamforming schemes improve with the number of passive reflecting elements, due to the increased beamforming gain.  Also, adding more semi-passive elements raises the performance of these beamforming schemes, especially in the case of a small number of passive reflecting elements.
It is worth noting that the performance improvement of the proposed beamforming scheme is due to both the increased beamforming gain and location sensing accuracy.

\subsection{Performance of the proposed ISAC transmission protocol}
Finally, we investigate the performance of the proposed ISAC transmission protocol, and define the average achievable rate as
{\color{black}
\begin{align}
   \bar{R}=\frac{1}{T_1+T_2} \sum\limits_{t=1}^{T_1+T_2} R (t).
\end{align}
}

\begin{figure}[!ht]
  \centering
  \includegraphics[width=4.5in]{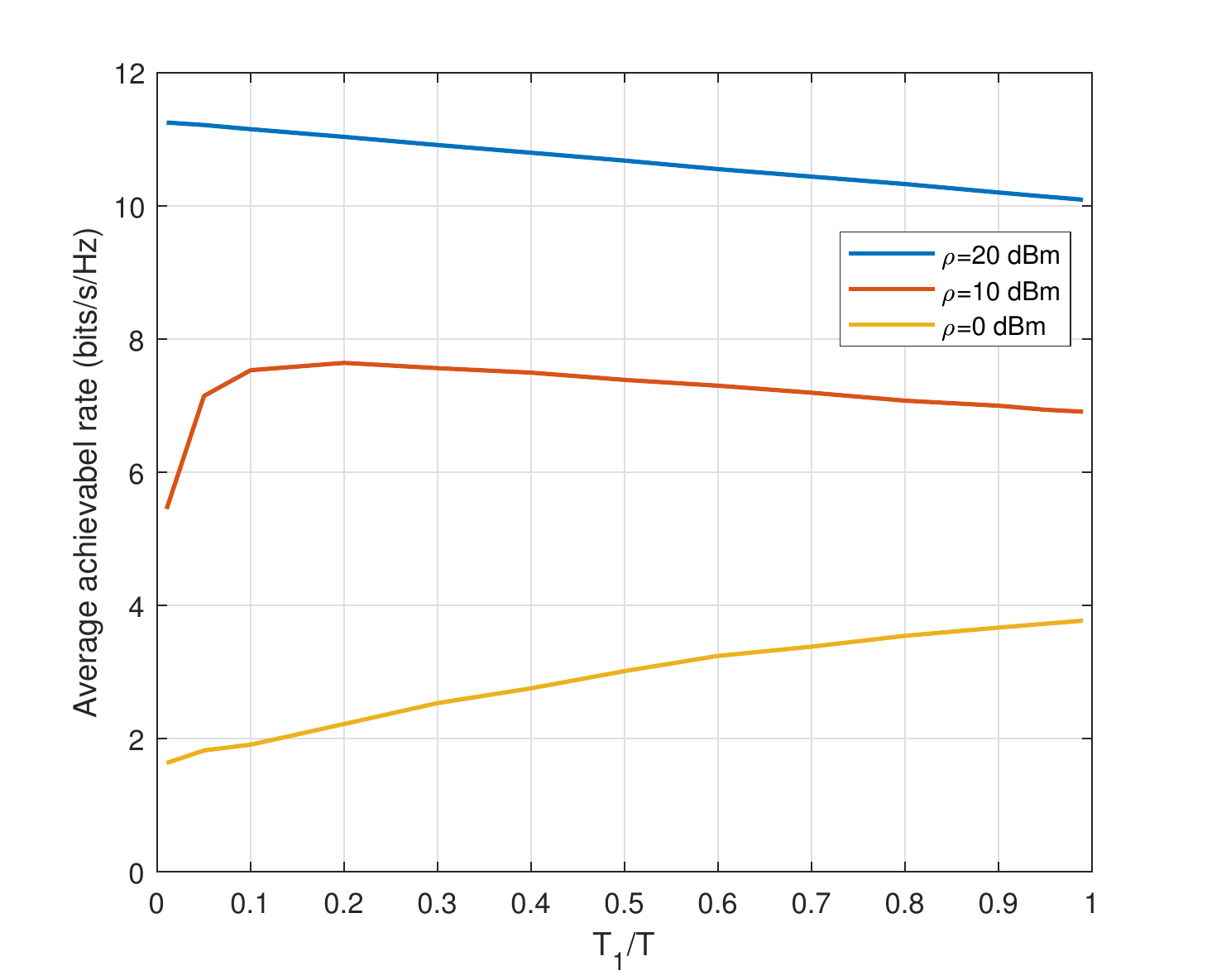}
  \caption{Performance of the proposed ISAC transmission protocol with different $T_1$/$T$.}
  \label{main7}
\end{figure}

Fig.\ref{main7} shows the average achievable rate of the proposed ISAC transmission protocol under different ratios of sensing time to the whole transmission time (i.e., $T_1$/$T$), where we set $\tau_1:T_1=\frac{1}{10}$ and $T=2000$.  The optimal ratio of sensing time to the whole transmission time decreases with the transmit power. For a low transmit power of $0$ dBm, the optimal ratio is about 1, which indicates that the two semi-passive sub-IRSs should always operate in the sensing mode. As the transmit power increases to $10$ dBm, the optimal ratio drops to 0.2.  With the highest transmit power of $20$ dBm, the optimal ratio approaches to 0. This is because the sensing accuracy with a short sensing time is enough high in this case and the two semi-passive sub-IRSs should operate in the reflecting mode to help data transmission.

\begin{figure}[!ht]
  \centering
  \includegraphics[width=4.5in]{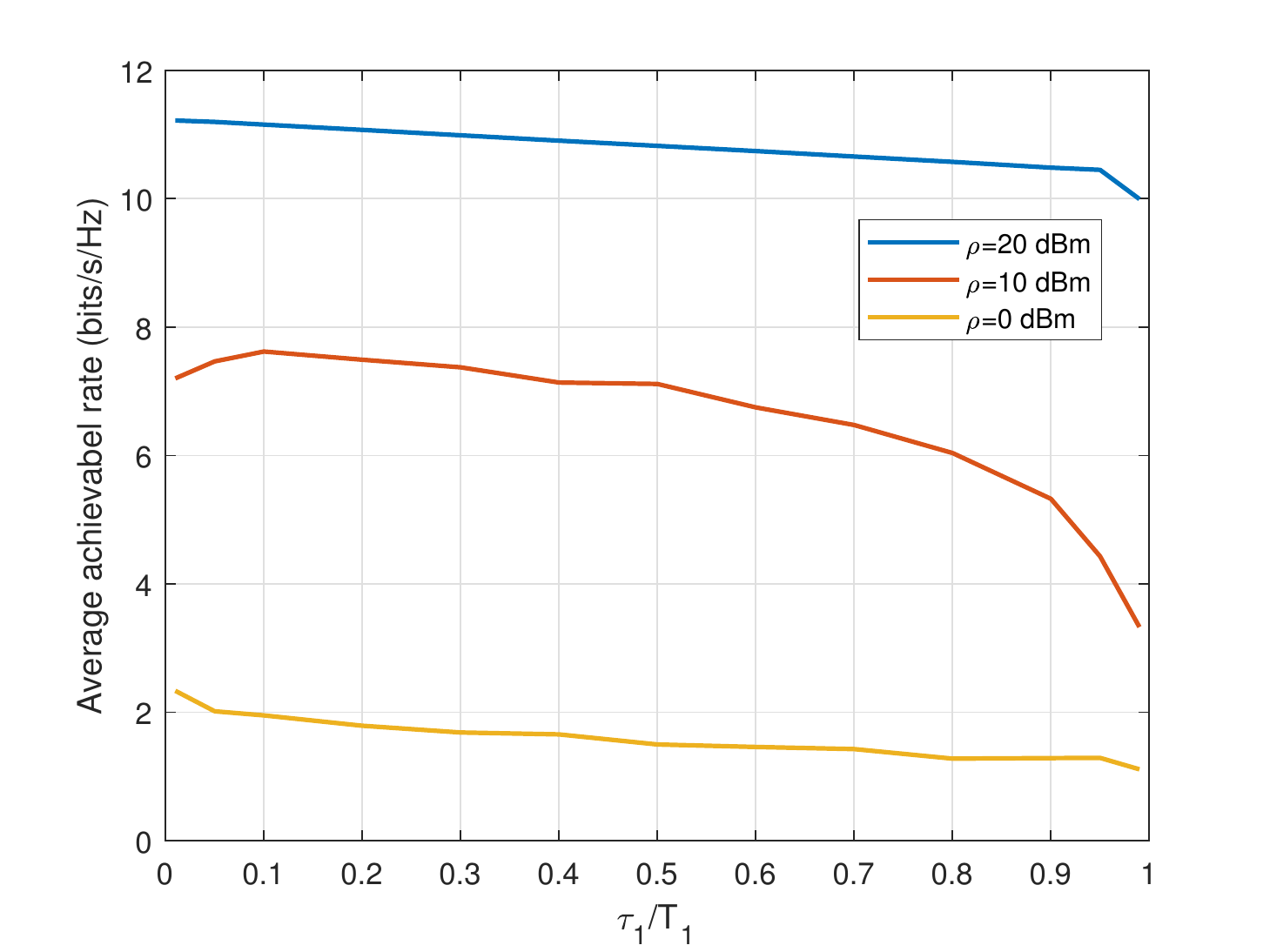}
  \caption{Performance of the proposed ISAC transmission protocol with different $\tau_1$/$T_1$.}
  \label{main8}
\end{figure}

Fig.\ref{main8} shows the average achievable rate of the proposed ISAC transmission protocol with different ratios of $\tau_1$ to $T_1$ (i.e., $\tau_1$/$T_1$), where we assume $T_1:T=\frac{1}{10}$ and $T=2000$. For all three configurations of transmit power, it is desired to allocate a very small portion of time slots to the first time block. This is because during the first time block, the BS does not have any CSI to design the IRS phase shifts and the corresponding achievable rate is very low. More time slots should be allocated to the second time block, where a much higher achievable rate can be achieved by properly designing  IRS phase shifts  according to the estimated user location during the first time block.

{\color{black}
Fig.\ref{main9} compares the proposed ISAC protocol with a benchmark protocol. For the benchmark protocol, location sensing and communication tasks are conducted, occupying different time resources. One coherence block consists of two periods. The first period is dedicated to pure location sensing with the assistance of the two semi-passive sub-IRSs, while the second period is dedicated to pure communication to the BS with the assistance of  all three sub-IRSs.
 As can be readily seen, the proposed ISAC protocol achieves its best performance when the time allocation ratio $T_1/T$  is $0.6$, while the benchmark protocol achieves its best performance when the time allocation ratio is $0.1$. Moreover, the optimal rate achieved by the proposed ISAC protocol is much higher than that achieved by the benchmark protocol. This is because our proposed ISAC protocol allows the location sensing task to be carried out without occupying any time resources of communication. 
  \begin{figure}[!ht]
  \centering
  \includegraphics[width=4.5in]{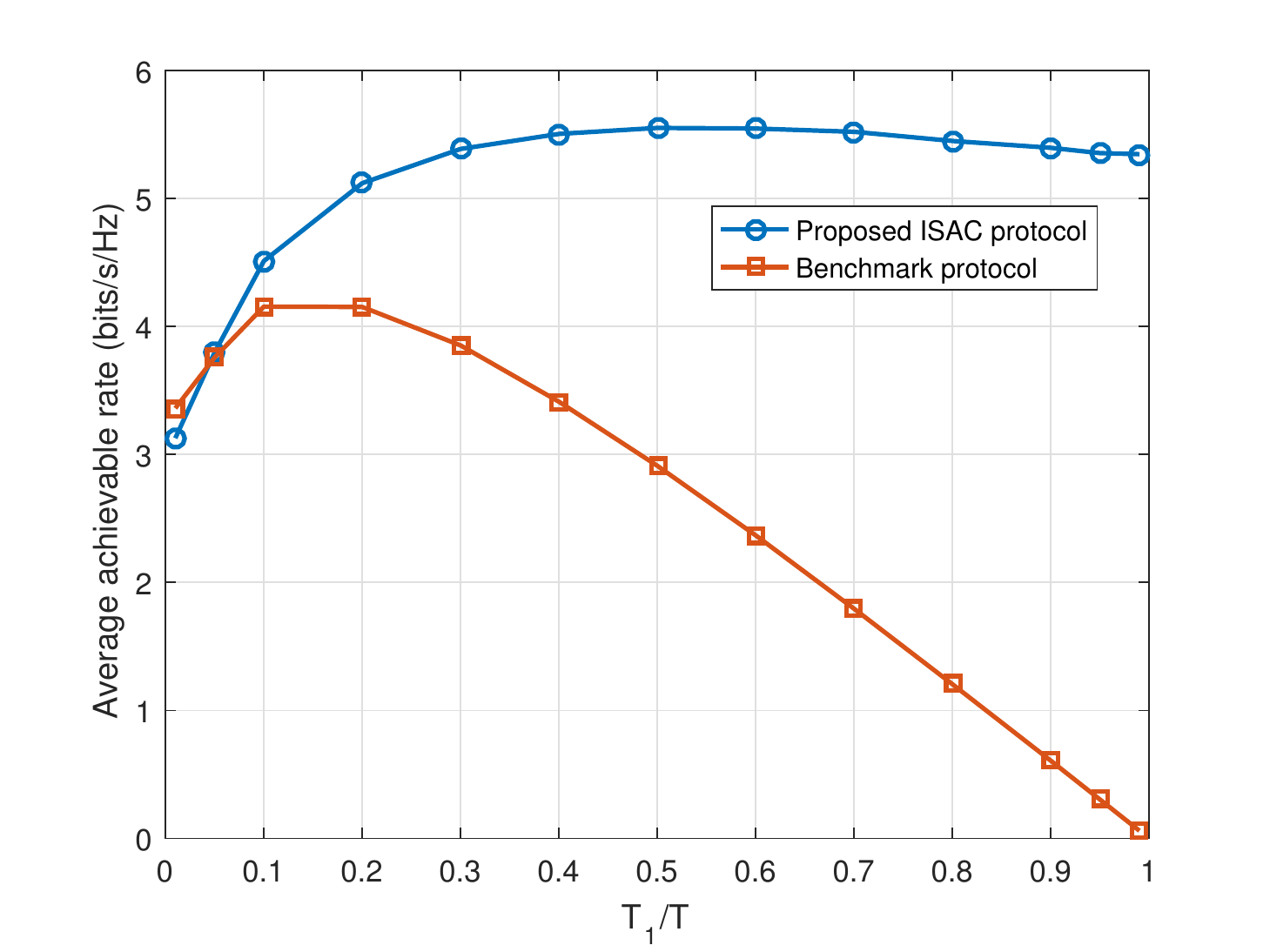}
  \caption{\color{black}Comparison between the proposed ISAC protocol and the benchmark protocol.}
  \label{main9}
\end{figure}
}

\section{conclusion} \label{s5}
In this paper, we construct an ISAC system by introducing a new IRS architecture (i.e., the distributed semi-passive IRS architecture) to the communication system. In the proposed IRS-based ISAC system, location sensing and data transmission can be conducted at the same time, occupying the same spectrum and time resources. The framework of the IRS-based ISAC system is designed, which includes the transmission protocol, location sensing and beamforming optimization.  Specifically, we consider a coherence block, where the user sends communication signals to the BS. The coherence block is composed of the ISAC period with two time blocks and the PC period. During the ISAC period, uplink data transmission with the assistance of the passive sub-IRS  and  location sensing at the two semi-passive sub-IRSs are conducted simultaneously.  The estimated location  in the first time block will be used for beamforming design in the second time block.  During the PC period, the whole distributed IRS operates in the reflecting mode for enhancing data transmission. The location obtained in the second time block of the ISAC period is used for phase shift design in the PC period.
Simulation results show that a millimeter-level positioning accuracy can be achieved with the proposed location sensing scheme. By increasing  the number of semi-passive elements or sensing time, the positioning accuracy can be further improved. Although only the imperfect location information is available, the proposed beamforming scheme for the ISAC period achieves almost the same performance as the optimal beamforming scheme with perfect CSI, and the proposed beamforming scheme for the PC period has similar performance to the AO beamforming scheme \cite{qingqing1} assuming perfect CSI. By investigating the trade-off between the sensing and communication performance, we find that increasing the sensing time will always improve the sensing performance, but  not always degrade the communication performance.


\bibliographystyle{IEEEtran}
\bibliography{references}{}

\end{document}